\documentclass[twocolumn]{aastex631}

\usepackage{color}
\usepackage{amsmath}

\begin{document}
\title{Effects of Dust in Electron-Capture Supernovae: Difficulty in Explaining Intermediate Luminosity Red Transients}
\author[0009-0006-7880-4959]{Yuma Takeichi}
\affiliation{Grinnell College, 1115 8th Ave, Grinnell, IA 50112, US}
\affiliation{Center for Astrophysics $|$ Harvard \& Smithsonian, 60 Garden St, Cambridge, MA 02138, USA}
\author[0000-0002-6347-3089]{Daichi Tsuna}
\affiliation{Center for Astrophysics $|$ Harvard \& Smithsonian, 60 Garden St, Cambridge, MA 02138, USA}
\author[0000-0002-1125-9187]{Daichi Hiramatsu}
\affiliation{Department of Astronomy, University of Florida, Bryant Space Science Center, Gainesville, FL 32611-2055, USA}
\author[0000-0002-1417-8024]{Morgan MacLeod}
\affiliation{Center for Astrophysics $|$ Harvard \& Smithsonian, 60 Garden St, Cambridge, MA 02138, USA}
\author[0009-0008-0189-863X]{Masato Sato}
\affiliation{Department of Liberal Arts, Tokyo University of Technology, 1404-1 Katakuramachi, Hachioji, Tokyo 192-0982, Japan}

\correspondingauthor{Yuma Takeichi}
\email{takeichi@grinnell.edu}

\begin{abstract}
The progenitors of intermediate-luminosity red transients (ILRTs) spanning the luminosity gap between novae and supernovae are under debate, and electron-capture supernovae (ECSNe) have been proposed as a strong candidate. ECSNe from super asymptotic giant branch (super-AGB) stars have been theoretically predicted to be much brighter than ILRTs, but attenuation by dusty winds of their super-AGB progenitors may ease this tension. Here we construct a semi-analytical framework for calculating the light curves of ECSNe with dust attenuation and re-emission, by combining recent ECSN light curve predictions with a model for dust physics. Even by varying both the wind mass-loss rate and the ECSN model parameters, we find it challenging to simultaneously fit the light curve and color evolution of ILRTs. This poses a potential challenge to the standard models of ECSNe for explaining ILRTs. For explaining ILRTs, we propose either the existence of a population of ECSNe with much lower explosion energies ($\sim$10$^{48}$--$10^{49}$ erg) and dimmer plateaus, or alternative model(s) to ECSNe that realize such low explosion energies.
\end{abstract}


\section{Introduction}
The development of optical surveys has led to the discovery of a population of ``gap transients" \citep{Kasliwal12,Pastorello2019,Cai2022}, with peak luminosities between those of classical novae and supernovae (SNe). These gap transients have been divided into different subtypes with likely different origins, based on their light curve features and spectral evolution. 

Among them, one of the most mysterious subclasses is intermediate luminosity red transients (ILRTs). Their light curves are characterized by a gradual increase in luminosity over weeks to peak optical magnitudes from $-11.5~\rm{mag}$ to $-14.5~\rm{mag}$, followed by a linear luminosity decline or plateau lasting 2-4 months \citep{Pastorello2019,Cai2021}. The late phase of the light curve has a slope comparable to what is powered by radioactive decay of $^{56}\rm Co$ (the daughter isotope of $^{56}$Ni), though some events have shallower slopes (e.g., AT2019abn, AT2019ahd, and NGC300-OT; \citealt{Valerin2025}). Another characteristic feature is the presence of narrow ($\lesssim1000$\,km\,s$^{-1}$) Balmer and [Ca II] features in their spectra \citep[e.g.,][]{Prieto2008-SN2008S, Bond2009-NGC,Cai2021,Valerin2025_spectra}, indicating existence of dense circumstellar matter (CSM).

Though the number of discovered ILRTs has increased over the past decades, their progenitors remain debated. ILRTs have been discussed as eruptions of luminous blue variables (LBVs; \citealt{Smith2009,Berger2009,2021Andrew}), post-red supergiant or post-asymptotic giant branch (AGB) stars \citep{Humphreys2011}, or an event driven by mass transfer from an accretion event of an extreme AGB star to its companion \citep{Kashi2010}. Besides the above scenarios of ILRTs being non-terminal explosions where their progenitors survive, other scenarios have been proposed where ILRTs are terminal explosions. These include failed SNe with dense CSM \citep{Tsuna2020}, weak explosions of blue supergiants \citep{Moriya2024}, and electron-capture SNe (ECSNe; \citealt{Thompson09,Pumo2009,Botticella2009,Kochanek2011,Cai2021,Rose2025}).

ECSNe are a subtype of core-collapse SNe (CCSNe) from super-AGB stars that occupy the low-mass end of CCSN progenitors \citep{1980Miyaji, Nomoto1984, Nomoto1987}. When the progenitor is not massive enough for the core to reach the temperature that can ignite neon, it forms a degenerate oxygen–neon–magnesium (ONeMg) core. When the core approaches the Chandrasekhar mass before the hydrogen and helium layers are completely stripped, the core, sustained by electron degeneracy pressure, collapses due to electron-capture reaction onto Mg and Ne nuclei. Once the central density reaches that of nuclear matter, the inner homologous core bounces. Though the energy of this shock is not enough to cause a prompt explosion, the shock can revive due to the small mass accretion rate and ram pressure as the edge of the ONeMg core has a steep density decline. 
First-principles simulations with neutrino transport have consistently found successful explosions with energies of $\sim (1$--$2)\times 10^{50}~ \rm erg$, which is weaker than canonical CCSNe \citep[e.g.,][]{2006Kitaura, 2008Janka, Radice17,Zha22}.
This would be a terminal explosion with the source eventually fading below the progenitor's luminosity, consistent with late-time observations of some ILRTs \citep{Adams2016,Valerin2025}.

\begin{figure*}[t]
    \centering    \includegraphics[width=0.9\textwidth]{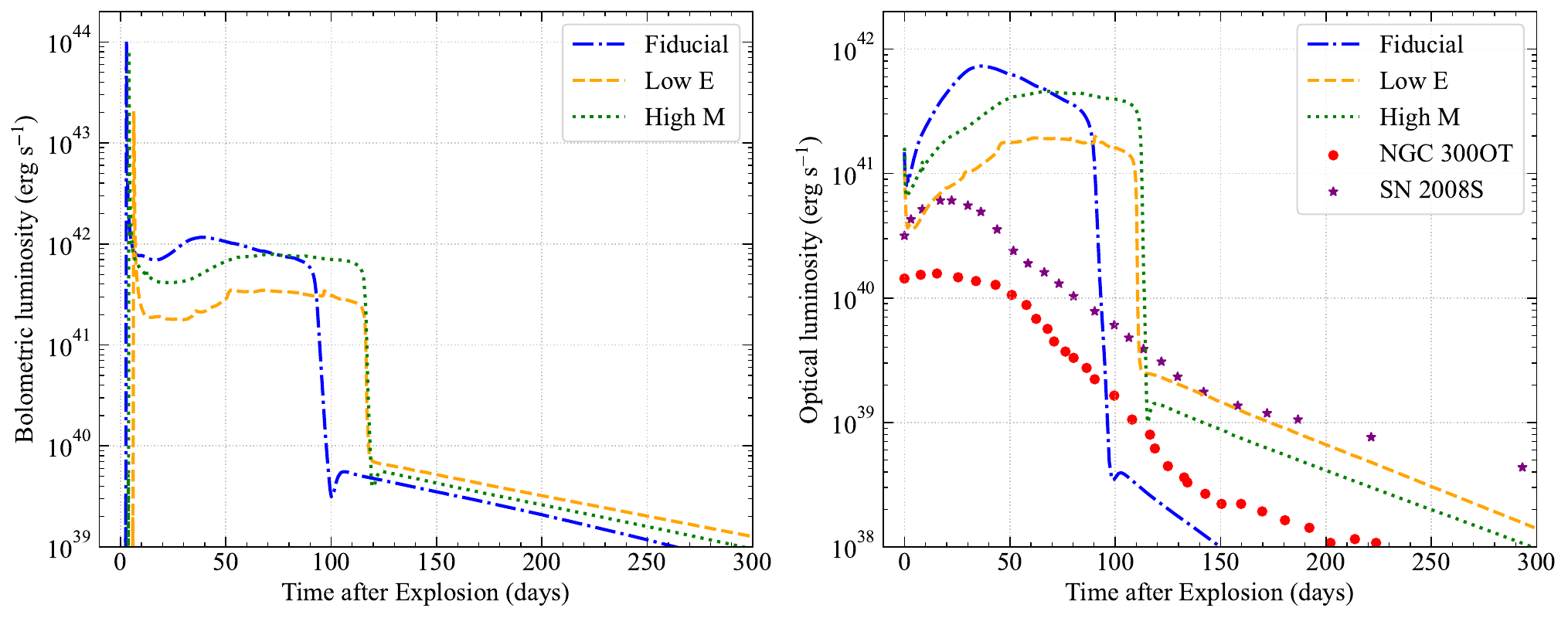}
    \caption{Light curves of three ECSN models from \citet{Sato2026} without CSM (see Table 
    \ref{table:model_info} for model parameters), with left and right panels respectively plotting the bolometric and optical ($B$ to $I/i$) luminosity. Data points on the right panel are optical light curves of two representative ILRTs with late-time coverage, NGC300-OT and SN 2008S (data from \citealt{Cai2021}; here time-shifted so that their first data point is at $t=0$ in this plot).}
    \label{fig:LCmodel}
\end{figure*}

However, ECSNe with explosion energies of $\sim 10^{50}~\rm erg$ result in peak optical magnitudes of $\lesssim-16$ mag, much brighter than ILRTs \citep{Tominaga13, Sato2024, Sato2026}. Furthermore from an observational perspective, SN 2018zd, one of the strongest candidates of ECSNe to date, is $10$--$100$ times brighter than ILRTs around peak \citep{2021Hiramatsu}.

A possibility to reconcile this apparent tension in optical brightness is dust extinction. Super-AGB stars could lose a significant part of their envelopes through dust-driven winds, with mass loss rates $\dot{M} \simeq (10^{-6}$--$10^{-3})~M_\odot~\rm yr^{-1}$ and velocities $v_w\simeq 10$--$25~\rm km~s^{-1}$ \citep[][and references therein]{2017Doherty}. Thus progenitors of ECSNe may be highly dust-enshrouded, which is indeed observed in pre-explosion data for some of the ILRTs \citep{Prieto2008-SN2008S,Thompson09,Kochanek2011,Jencson19}. A detailed comparison combining theoretical ECSN light curves with dust physics could put this scenario into a quantitative test.

In this work, we construct a semi-analytical framework for calculating the light curves after attenuation and re-emission due to dust. We combine a model for destruction and re-formation of circumstellar dust based on an earlier study of \cite{Kochanek2011}, with the recent light curve models of ECSNe from \cite{Sato2026}. By comparing the resulting light curves to long-term observations of representative events (NGC300-OT and SN 2008S), we test the possibility of ECSNe attenuated by dust as an explanation for ILRTs. 

This paper is structured as follows. In Section \ref{sec:ECSNe model by Sato+26}, we introduce the CSM-free ECSN light curve models by \citet{Sato2026}. In Section \ref{sec:model}, we construct our semi-analytical model including CSM interaction and dust physics. In Section \ref{sec:result}, we compare our models with dust attenuation to ILRTs, and find that for reasonable variations of the CSM density structure, it is difficult to simultaneously reproduce the overall evolution of luminosity and color with dust.

In Section \ref{sec:summary and discussion}, we summarize our results and argue that this may suggest an intrinsically dim explosion for ILRTs. This poses a potential challenge for existing ECSN scenarios, and we discuss possible alternative scenarios for the progenitors of ILRTs.

\section{ECSN Light Curve models}
\label{sec:ECSNe model by Sato+26}
We adopt the light curve models of CSM-free ECSNe recently calculated by \cite{Sato2026}. This work explodes models of super-AGB stars, using the multi-group radiation hydrodynamical code STELLA \citep{1993Blinnikov, 1998Blinnikov, 2000Blinnikov}. The super-AGB progenitors are constructed by attaching hydrogen-rich envelopes of various masses to a 1.377M$_\odot$ ONeMg core from \cite{Nomoto1984, Nomoto1987} under hydrostatic and thermal equilibrium \citep{Tominaga13}. The distribution of explosively synthesized elements, including $\sim2\times10^{-3}~M_\odot$ of radioactive $^{56}$Ni, is adopted from \cite{Wanajo2009}.

We adopt three models in Table \ref{table:model_info} spanning a range of ejecta mass $M_{\rm ej}$ and explosion energy $E_{\rm ej}$, representing the expected range of light curve properties in ECSNe. Figure \ref{fig:LCmodel} shows the light curves of these models, with the left and right panels respectively being the bolometric luminosity and the optical luminosity integrated over the wavelength range 430--890~nm ($B$ to $I$ band). 
The light curve models all have three key features: shock breakout in the first days with peak bolometric luminosities of $L_{\rm sb}\approx 10^{43}$--$10^{44}$ erg s$^{-1}$, plateau lasting $\sim$ 100 days with luminosities of $10^{41}$--$10^{42}$  erg s$^{-1}$ mainly emitted in optical, and a tail in the late phase of $\lesssim 10^{40}$ erg s$^{-1}$ powered by $^{56}$Co decay. 

To illustrate the discrepancy between models and observed ILRTs, in the right panel we compare these models to the light curves of two representative ILRTs, SN 2008S and NGC300-OT. The theoretical light curves of ECSNe are very distinct from ILRTs in two ways. First the plateau luminosity of ECSNe is much higher than those of ILRTs ($\sim10^{40}$--$10^{41}$ erg s$^{-1}$ at around the same epochs). Second and more importantly, the ECSN models display a drastic drop in luminosity when transitioning to the $^{56}$Co tail, in contrast to ILRTs that have a much smoother light curve evolution. This sharp drop is due to the combination of the bright plateau phase owing to the large progenitor radius and the dim tail phase from the small amount of synthesized $^{56}$Ni.

\begin{table*}
\centering

\begin{tabular}{ccccc|ccc}
\hline
Model & $M_{ej}$ & $E_{ej}$ & $R$ & $M_{\rm Ni}$ & $L_{\rm sb}$ & $L_{\rm opt,50}$ & $t_{\rm pl}$\\
& ($M_\odot$)  & ($10^{50}$ erg) & ($R_\odot$) & ($10^{-3} M_\odot$) & ($10^{43}$ erg s $^{-1}$) & ($10^{41}$ erg s $^{-1}$) & (day) \\
\hline
\hline
Fiducial & 3.17 & 1.02 & 1018 & 1.62 & 10.4 & 6.30 & 91.2\\
Low E  & 3.16 & 0.196 & 1000 & 1.67& 2.06 &1.78 & 110\\
High M  & 4.82 & 0.949 & 1049 &1.63& 8.24 & 4.06 & 113\\
\hline
\end{tabular}
\caption{
Light curve characteristics of ECSN models from \cite{Sato2026} adopted in this work. $M_{ej}$, $E_{ej}$, $R$, $M_{\rm Ni}$ are respectively the ejecta mass, explosion energy, radius of the progenitor envelope (constructed under hydrostatic and thermal equilibrium), and mass of $^{56}$Ni in the ejecta in each model. 
$L_{\rm sb}$ is the shock breakout luminosity calculated as the maximum of the bolometric luminosity, $L_{\rm opt, 50}$ denotes the optical luminosity at the plateau phase (here measured at 50 days after the explosion), and $t_{\rm pl}$ is the end of the plateau phase defined here as the minimum of the (log-)luminosity gradient $d\log L/dt$.
}
\label{table:model_info}
\end{table*}

\section{Including a Dusty Wind}
\label{sec:model}
\begin{figure*}[th]
    \centering    \includegraphics[width=\linewidth]{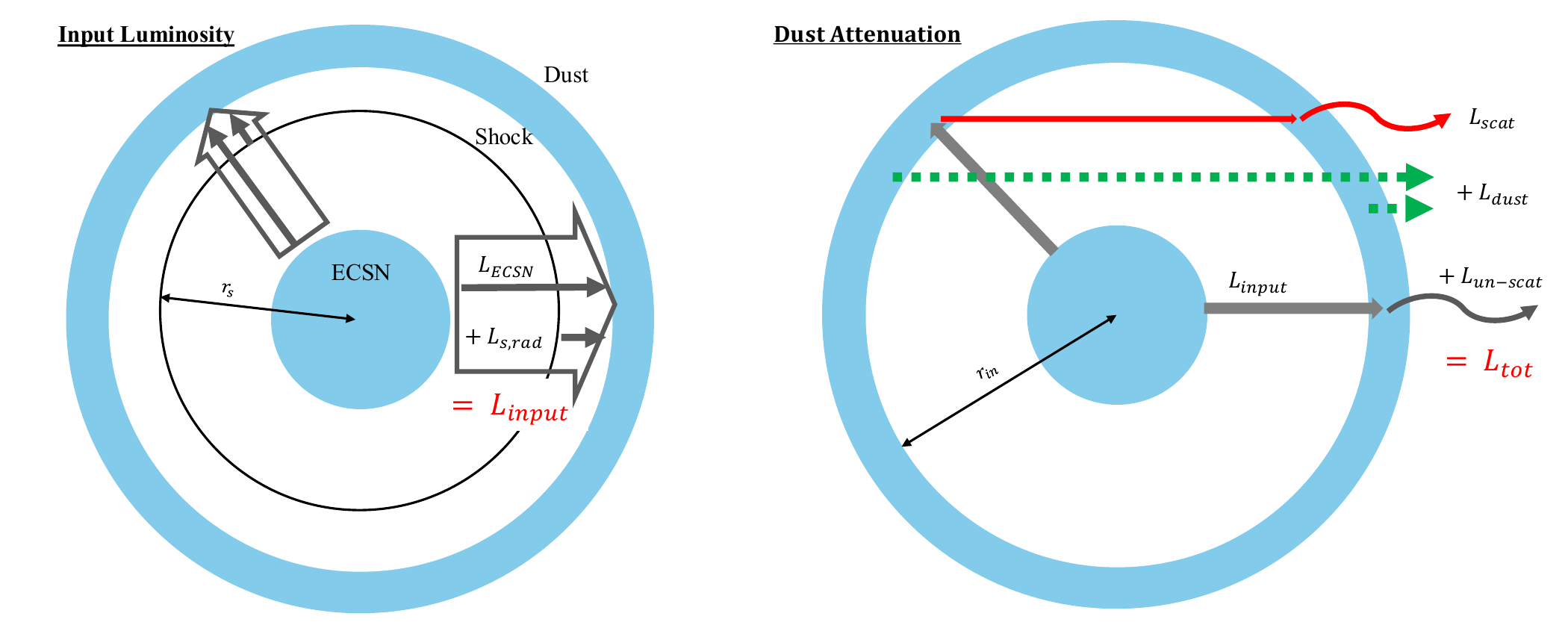}
    \caption{A schematic of our model for calculating light curves of ECSNe attenuated by dust. Left figure shows our calculation of the input spectral energy distribution (SED) $L_{\rm input}$, consisting of the ECSN luminosity ($L_{\rm ECSN}$; Section \ref{sec:ECSNe model by Sato+26}) and luminosity from the interaction between the SN ejecta and the wind ($L_{\rm s, rad}$; Section \ref{subsec: Input Luminosity}). Right figure shows the calculation of the final observed luminosity after dust attenuation (Section \ref{subsec:LC_with_dust}), consisting of three components: escaping luminosity after attenuation without scattering by dust ($L_{\rm un-scat}$), escaping luminosity with scattering ($L_{\rm scat}$), and dust re-emission ($L_{\rm dust}$). The evolution of the dust layer and its optical depth are described in Sections \ref{subsec:dustphysics} and \ref{subsubsec: opacity} respectively.}
    \label{fig:modelimage}
\end{figure*}

The CSM-free models above only calculate the emission diffusing out from the SN ejecta. For the expected wind mass-loss rates, interaction between the ECSN ejecta and CSM could significantly contribute to the luminosity, especially after the plateau phase \citep[e.g.,][]{Moriya14}. Furthermore, the high mass-loss rates of these winds can lead to strong dust attenuation. Our goal is to semi-analytically include these effects for more concrete comparisons to ILRTs. 

Figure \ref{fig:modelimage} schematically shows our model for calculating the light curves including dust extinction. We calculate the input luminosity $L_{\rm input}$ (left panel) arising from two components, the ECSN itself (Section \ref{sec:ECSNe model by Sato+26}) and the interaction between the ECSN ejecta and super-AGB progenitor's wind. We then model the attenuation, re-emission, and scattering due to dust before reaching the observer (right panel). Our modeling for the former input luminosity and the latter dust physics is s described in Section \ref{subsec: Input Luminosity} and Sections \ref{subsec:dustphysics}--\ref{subsec:LC_with_dust}.

As shown in the right panel of Figure \ref{fig:modelimage}, the final observed light curve is composed of three components. First, $L_{\rm unscat}$ corresponds to the component that is not scattered by dust and is transmitted following dust attenuation.
Second, $L_{\rm scat}$ corresponds to the component that is scattered by dust and is transmitted following dust attenuation. Finally, $L_{\rm dust}$ corresponds to the luminosity from dust re-emission. The process for modeling the optical luminosity from each of these components is described in Section \ref{subsec:LC_with_dust}.

\subsection{Including luminosity from CSM interaction}
\label{subsec: Input Luminosity}
We assume spherically symmetric SN ejecta and wind CSM, with the latter having a mass-loss rate $\dot{M}$ and terminal velocity fixed hereafter as $v_w = 10~ \text{km} \ \text{s}^{-1}$ characteristic of (super-)AGB stars. The radial density profile of the wind well outside the progenitor is given by that of a stationary wind as 
\begin{eqnarray}
    \rho_{\rm wind}(r) = qr^{-2},
\end{eqnarray}
where
\begin{eqnarray}
    q&=& \frac{\dot{M}}{4\pi v_w} \approx 5.0\times 10^{14}\ {\rm g\ cm^{-1}} \dot{M}_{-4} \left(\frac{v_w}{10\ {\rm km\ s^{-1}}}\right)^{\!\!-1}.
    \label{eq: wind density}
\end{eqnarray}
Hereafter we use $\dot{M}_{-4} = \dot{M}/(10^{-4}~ M_\odot~{\rm yr}^{-1})$ to scale the mass-loss rate by its value expected from theoretical predictions and existing wind prescriptions \citep[e.g.,][]{Poelarends08,2017Doherty}. We adopt a double power-law density structure for the homologously expanding ECSN ejecta \citep{Matzner1999}
\begin{eqnarray}
\rho_{\rm ej} (v_{\rm ej},t)
&=& \left\{ \begin{array}{ll}
t^{-3}\left(v_{\rm ej}/g\right)^{-n} & (v_{\rm ej} > v_t),\\
t^{-3}(v_t/g)^{-n} \left(v_{\rm ej}/v_t\right)^{-\delta}  & (v_{\rm ej} < v_t),
\end{array}\right.
\label{eq:rho_ej}
\end{eqnarray}
which is valid after the ejecta expand to a few times the stellar radius and kinetic energy dominates over thermal energy. Constants $g$ and $v_t$ are functions of the ejecta mass $M_{\rm ej}$ and explosion energy $E_{\rm ej}$,
\begin{eqnarray}\label{eq:coeff_ej}
g &=& \left\{\frac{1}{4\pi(n-\delta)} \frac{[2(5-\delta)(n-5)E_{\rm ej}]^{(n-3)/2}}{[(3-\delta)(n-3)M_{\rm ej}]^{(n-5)/2}}\right\}^{1/n},\\
v_t &=& \left[\frac{2(5-\delta)(n-5)E_{\rm ej}}{(3-\delta)(n-3)M_{\rm ej}}\right]^{1/2}.\label{eq:v_t}
\end{eqnarray}

Since the super-AGB progenitors of ECSNe have convective envelopes which are similar to red supergiants, we adopt $n = 12$ and $\delta = 0.5$ as suggested in previous works \citep{Matzner1999,Moriya2013}. 

Over the observed timescale of $\lesssim$ year and for the expected wind mass-loss rates of $\dot{M}\lesssim 10^{-3}~M_\odot\ {\rm yr}^{-1}$, only the outer part of the ejecta (with $\rho_{\rm ej}\propto v_{\rm ej}^{-n}$) will participate in the CSM interaction. We thus calculate the radial evolution of the shock using the self-similar solution from \cite{Chevalier1982},
\begin{eqnarray}
    r_s(t) &\simeq& (Ag^n/q)^{1/(n-s)}t^{(n-3)/(n-s)}\nonumber  \\
    &\sim& 2 \times 10^{15} {\rm cm} \left(\frac{q}{10^{15}~{\rm g\ cm^{-1}}}\right)^{\!\!-1/10}\left( \frac{E_{\rm ej}}{10^{50}~\rm erg} \right)^{\!\!9/20}\nonumber\\ 
    &&\left(\frac{M_{\rm ej}}{5~M_\odot}\right)^{\!\!-7/20} \left(\frac{t}{100~{\rm day}}\right)^{\!\!9/10}\ (n=12, s=2),
    \label{eq:r_sh_scaling}
\end{eqnarray}
where $s = 2$ for our case of a stationary wind, and $A$ is a dimensionless constant that is $\simeq 0.038$ for $(n,s)=(12,2)$ \citep[see Table 1 of][]{Chevalier1982}.

As the shock velocity $v_s=dr_s/dt$ is typically of the order of (a few) $\times 10^3$ km s$^{-1}$ and always much larger than $v_w$, the shock dissipates the kinetic energy of the ejecta with luminosity 
\begin{equation}
    L_s(t) \simeq \frac{\dot{M}}{2}\frac{v_s^3}{v_w} \propto t^{-0.3}~(n=12, s=2),
    \label{eq:L_shock}
\end{equation}
and heats the CSM to a post-shock temperature
\begin{equation}
    T_s  \simeq \frac{3\mu}{16}\frac{m_p v_s^2}{k_B} \sim 1\times 10^8\ {\rm K}\left(\frac{v_s}{3000\ {\rm km\ s^{-1}}}\right)^2.
\end{equation}
Here $\mu\approx 0.6$ is the mean molecular weight for near-solar abundance, $m_p$ is the proton mass, and $k_B$ is the Boltzmann constant. 

We follow \cite{Kochanek2011} to calculate the radiated luminosity from shock interaction, by calculating the cooling timescale of shock-heated gas

\begin{align}
    t_{\rm{cool}} &= \frac{3nk_BT_s}{2\Lambda(n,T_s)} = \frac{3 k_BT_s\dot{M}}{2 \pi \Lambda \mu m_pr_s^2 v_w}
\end{align}
where $n = \dot{M}/({\pi \mu m_p r_s^2v_w})$ is the number density of shock-compressed gas for a compression ratio of 4, and
\begin{eqnarray}
    \Lambda(n,T) = A_0n^2\sqrt{T/T_0}
\end{eqnarray}
is the cooling rate assuming free-free emission is dominant, with $A_0 \simeq 2\times10^{-24} \rm{erg \ cm^3 \ s^{-1}}$ at $T=T_0 = 10^7~{\rm K}$. 
The radiated luminosity is then expressed as
\begin{equation}
    L_{s, \rm{rad}}(t) = L_s(t) \times {\rm min}\Big(1, \frac{t_{\rm{exp}}}{t_{\rm{cool}}}\Big) \label{eq: Lsrad}
\end{equation}
where $t_{\rm exp}=r_s/v_s$ is the dynamical timescale.
\begin{figure}
    \centering
    \includegraphics[scale=0.45]{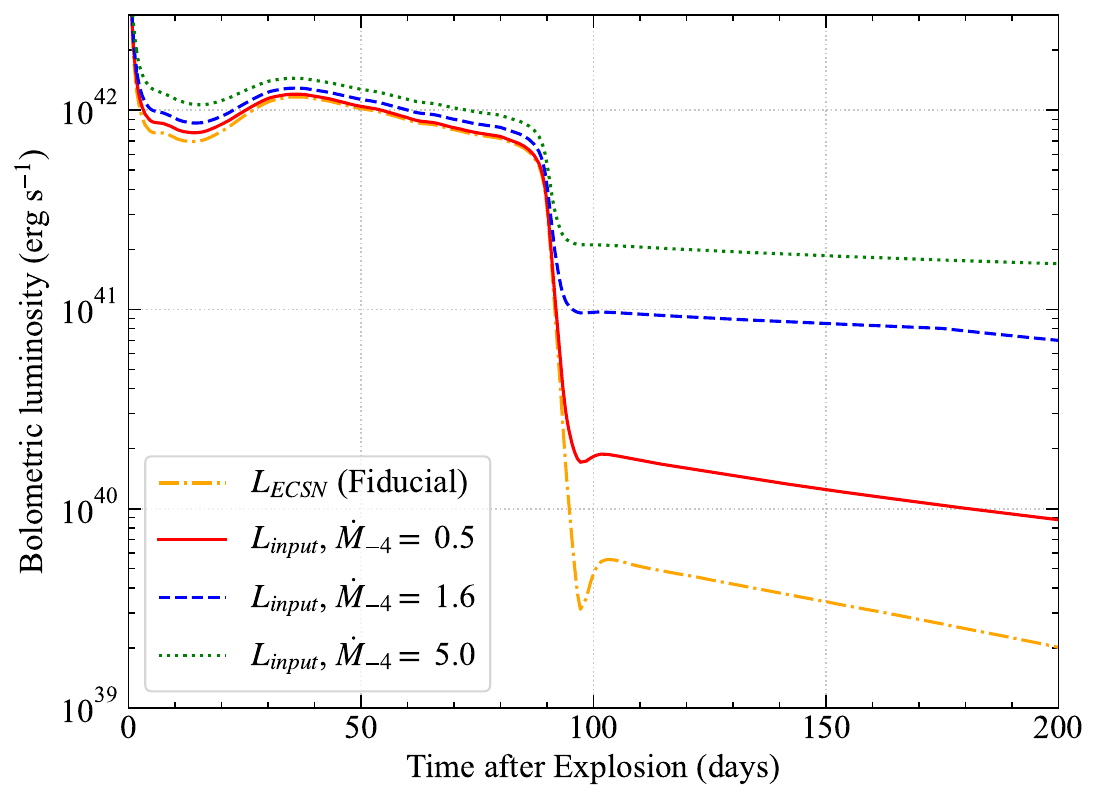}
    \caption{Evolution of the bolometric input luminosity $L_{\rm input}$ for the Fiducial ECSN model in Table \ref{table:model_info}, with the luminosity from shock interaction $L_{\rm s, rad}$ added for different wind mass loss rates of $\dot{M}_{-4} = [0.5,1.6,5]$ (where $\dot{M}_{-4} = \dot{M}/10^{-4}~ M_\odot ~ \rm yr^{-1})$.}
    \label{fig:analytical Ls}
\end{figure}
We finally obtain the bolometric input luminosity $L_{\rm input}$ for a given $\dot{M}$ as
\begin{equation}
    L_{\rm input} = L_{\rm ECSN} + L_{s, \rm rad},
\end{equation}
where $L_{\rm ECSN}$ is the bolometric luminosity of the ECSN component in Section \ref{sec:ECSNe model by Sato+26}. 

Figure \ref{fig:analytical Ls} shows how $L_{\rm input}$ changes for different mass loss rates over a range of $\dot{M}_{-4} = 0.5$--$5$, for the Fiducial ECSN model in Table \ref{table:model_info}. While the plateau is largely unchanged, the tail phase becomes much more luminous due to CSM interaction, with luminosity at the beginning of the tail phase being $\approx (0.2-2)\times10^{41}$ erg s$^{-1}$. In the tail phase, we see the tail has a steeper slope for lower $\dot{M}_{-4}$.
Noting $t_{\rm cool}\propto T_s^{1/2}/n\propto r_s^2v_s/\dot{M}$,
\begin{equation}
    \dfrac{t_{\exp}}{t_{\rm cool}}\propto \dfrac{\dot{M}}{r_sv_s^2}\propto\dot{M}^{1.3} t^{-0.7}.
\end{equation}
When $\dot{M}_{-4}=1.6, 5.0$, $t_{\rm cool} < t_{\rm exp}$ and $L_{s, \rm rad} = L_s$. On the other hand, for a lower mass loss rate ($\dot{M}_{-4} = 0.5$), $t_{\rm cool} > t_{\rm exp}$ and the time-dependent factor ${t_{\exp}}/{t_{\rm cool}}$ makes the slope steeper.

As expected and consistent with more detailed numerical calculations of \cite{Moriya14}, including the shock power $L_{s,\rm rad}$ alleviates the sharp light curve drop after the plateau depicted in Figure \ref{fig:LCmodel}. However, even for an extreme mass loss of $\dot{M} = 5\times 10^{-4} ~M_\odot~{\rm yr}^{-1}$, we still find a significant drop in the light curve after the plateau by a factor $\sim 5$. 
Therefore, modifying the ECSN model for ILRTs via dust requires, at a minimum, simultaneously suppressing both the bright plateau and the sharp luminosity drop following the plateau.

\subsection{Dust Formation and Growth}
\label{subsec:dustphysics}

The shock breakout of the ECSN will initially destroy the dust in the super-AGB wind out to a large radii. After the shock breakout quickly fades within days, dust will gradually reform as the gas recombines and cools \citep[][see their Figure 11]{Kochanek2011}. 

We construct a model for dust formation and growth, mainly following the physics described in \cite{Kochanek2011}. 

We impose two conditions for dust formation, that the dust should (i) not be sublimated by radiation from the subsequent SN light, and (ii) not be within the shock radius $r_s$ (equation \ref{eq:r_sh_scaling}), as shock heating leads to dust destruction via kinetic sputtering. The former is set by the characteristic sublimation radius
\begin{eqnarray}
    r_{\rm sub}(t) &=& \sqrt{\frac{L_{\rm input}(t)}{16\pi\sigma T_{\rm dest}^4Q_{\rm rat}}} \nonumber \\
    &\sim& 8\times 10^{15}\ {\rm cm} \left(\frac{L_{\rm input}(t)}{10^{42}\ {\rm erg\ s^{-1}}}\right)^{\!\!1/2},
    \label{eq: rsub_scaling}
\end{eqnarray}
where $\sigma$ is the Stefan-Boltzmann constant, $T_{\rm dest}$ is the dust destruction temperature set as $T_{\rm dest} = 1500~\text{K}$, and $Q_{\rm rat}$ is the ratio of absorption efficiency set as $Q_{\rm rat} = 1$ following \cite{Kochanek2011}. The two conditions 
allow dust to only form outside $r_{\rm in}(t)={\rm max}(r_{\rm sub},r_s)$. For each radius, we define $t_{\rm form}(r)$ as the time when the condition for dust reformation is first satisfied.

Once dust is allowed to reform from at a given radii, we can follow the growth of its grain size $a$. For a carbon-rich wind, the growth rate by carbon atoms sticking on the grain is approximately given as \citep{Kochanek2011}
\begin{eqnarray}
    \frac{da}{dt} &\approx& \frac{1}{4\pi a^2\rho_{\rm bulk}} (X_c\rho_{\rm wind})(\pi a^2)v_c \nonumber \\
    &\approx& \frac{X_c v_c}{4\rho_{\rm bulk}}\frac{\dot{M}}{4\pi r^2v_w},
    \label{eq:dadt}
\end{eqnarray}
where $X_c\sim 10^{-2}$ is the carbon mass fraction expected for carbon-enhanced solar metallciity gas, and $v_c\approx v_w = 10~\text{km}~\text{s}^{-1}$ is the effective collision velocity. Hereafter, we assume that the chemical composition of the dust is amorphous carbon, based on recent infrared (IR) observations finding strong features of carbonaceous dust typical of carbon-rich AGB stars in the ILRT AT2019abn \citep[][see also \citealt{Prieto09,Wesson10}]{Rose2025}. We thus adopt a bulk density $\rho_{\rm bulk} \simeq 2.0~ \text{g}~\text{cm}^{-3}$ \citep[][see their Table 2]{1991Martin}. 

Even if the above condition on dust formation is met at a given radii, build-up of dust is gradual. \cite{Kochanek2011} identifies the bottlenecks to be the recombination of gas ionized by the shock breakout, and the growth of the dust grains. The recombination rate $\alpha_B\rho_{\rm wind}/(\mu m_p)$ (where $\alpha_B\approx 10^{-13}$ cm$^3$ s$^{-1}$ is the Case-B recombination coefficient of hydrogen)  and the dust growth rate (equation \ref{eq:dadt}) respectively give the limiting timescales $t_{\rm{rec}}$ and $t_{\rm{grow, amin}}$ for these two processes
\begin{align}
   t_{\rm{rec}}(r) &= \frac{\mu m_p}{\alpha_B \rho_{\rm wind}} \nonumber \\ 
                &\approx 23~{\rm days}~\dot{M}_{-4}^{-1}\left(\frac{v_w}{10~\rm{km} ~s^{-1}}\right)\left( \dfrac{r}{10^{16}~\rm cm}\right)^{2} \label{eq:t_rec}\\
t_{\rm{grow, amin}} & =a_{\rm{min}}\dfrac{16 \pi \rho_{\rm{bulk}}v_wr^2}{X_c v_c\dot{M}} \nonumber\\
&\approx 55~{\rm days}~\dot{M}_{-4}^{-1}\Big(\dfrac{v_c}{v_w}\Big)^{\!\!-1}
\Big( \dfrac{r}{10^{16}~\rm cm}\Big)^{2} \Big( \dfrac{a_{\rm{min}}}{0.3~\rm nm}\Big)
\label{eq:tgrow_amin}
\end{align}
where we define $a_{\rm{min}} = 0.3 ~ \rm{nm}$ as the minimal size for the carbonaceous grains/particles to exist \citep[e.g.,][]{Omont86,Guhathakurta89}.
These scalings imply three key points. First, dust does not form immediately but with a timescale of 10--100 days depending on $\dot{M}$, and the evolution of dust optical depth would thus be quite sensitive to $\dot{M}$. Second, as these timescales have a strong radial dependence ($\propto r^{2}$), we expect most of the dust re-formation to first occur in the vicinity of the innermost radii $r_{\rm in}$. Third, dust would have a range of grain size that varies with radius, with larger grains at inner radii.

To capture this gradual and radial-dependent formation and growth of dust over time, we introduce a variable $f_{\rm{dust}}(r,t)$ that sets the fraction of the wind material that could form dust grains
\begin{equation}
f_{\rm{dust}}(r,t) =  \frac{\Delta t}{\Delta t+{\rm max}(t_{\rm{rec}}, t_{\rm{grow, amin}})}\,
\end{equation}
where $\Delta t = t - t_{\rm{form}}(r)$ is the elapsed time from dust formation. 
This equation smoothly captures the linear rise of $f_{\rm dust}$ when $\Delta t \ll \max (t_{\rm{rec}}, t_{\rm{grow, amin}})$, and the saturation to unity when $\Delta t \gg \max (t_{\rm{rec}}, t_{\rm{grow, amin}})$.

\subsection{Dust Optical Depth Evolution}
\label{subsubsec: opacity}
Given the dust formation fraction $f_{\rm dust}$ defined above, the dust optical depth at a given wavelength $\lambda$ is
\begin{equation}
\tau_{\lambda}(t) \simeq \int^{\infty}_{r_{\rm in}} \kappa_{\lambda}(r,t) [f_{\rm{dust}}(r,t)\rho_{\rm wind}(r)]dr
\end{equation}
where $\kappa_\lambda$ is the wavelength-dependent opacity that is set by the maximal grain size $a_{\rm max}(r,t)\approx a_{\rm min}+(da/dt)(t-t_{\rm form})$ attained by dust growth. 
Here the upper limit of the integral is conveniently set to infinity, as the rate of dust re-formation would be negligible at increasingly larger radii (equations \ref{eq:t_rec} and \ref{eq:tgrow_amin}). However, we note that at very large radii there could be contribution from {\it pre-existing} dust that survives the shock breakout. We later discuss the effects of surviving dust in Section \ref{sec:surviving_dust}.

\begin{figure}[!t]
    \centering
    \includegraphics[scale=0.45]{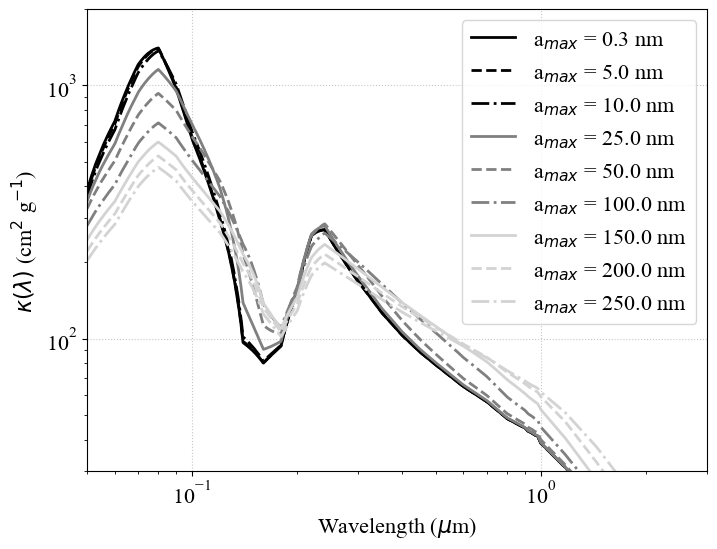}
    \caption{Wavelength-dependent dust opacity, for a MRN-like ($\propto a^{-3.5}$) distribution with varying $a_{\rm{max}} = 0.3$--$250~{\rm nm}$ and composition of amorphous carbon. For all cases, we fix $a_{\rm min}=0.3$ nm.}
    \label{fig:opacity&amax}
\end{figure}

To obtain $\kappa_\lambda$, we use the outputs from the dust radiative transfer code DUSTY \citep{Ivezic97,Ivezic99}. We solve the radiative transfer in a spherical dusty wind, with a density structure of an analytical profile for a radiatively-driven wind ("RDW-analytic") and with a relative shell thickness of $Y=10^3$ from the initial (dust condensation) radius $r_1$ to the outer edge. We assume the composition of the dust to be amorphous carbon (\texttt{amC-Hn} in DUSTY), and a grain size distribution of a modified MRN distribution \citep{1977Mathis} with varied $a_{\rm max}$
\begin{equation}
    n(a) \propto a^{-q} \ \text{for} \  a_{\rm min} < a < a_{\rm max},
\end{equation}
where we fix $q = 3.5$, $a_{\rm min} = 0.3~ \text{nm}$.\footnote{While equation (\ref{eq:dadt}) captures the characteristic growth of dust grains, it does not reliably predict the more complicated evolution of the grain-size distribution. We set $q = 3.5$ for simplicity motivated from the MRN distribution, but we verified that variation of $q$ does not change the opacity drastically. 
} We adopt a grid of 1664 wavelength bins in the range $0.04$--$905~ \mu \text{m}$, for which the optical constants are available in DUSTY. 
For given inputs $\tau_V$ and $a_{\rm max}$, DUSTY outputs the wavelength-dependent optical depth $\tau_{\lambda, \rm DUSTY}$. From this, $\kappa_{\lambda}$ follows
\begin{align}
    \kappa_\lambda &= \dfrac{\tau_{\lambda, \rm DUSTY}}{\displaystyle \int^{Yr_1}_{r_1} \frac{\dot{M}_{\rm DUSTY}}{4\pi r^2v_{\rm DUSTY}(r)}dr} .
\end{align}
In this ``RDW-analytic" case, DUSTY also outputs the corresponding mass loss rate $\dot{M}_{\rm DUSTY}$ under an assumed dust-to-gas ratio of $200$ expected for solar metallicity, and terminal velocity under the wind acceleration modeled by
\begin{eqnarray}
    v_{\rm DUSTY}(r) = v_e \times \sqrt{1-\frac{r_1}{r}\left[1-\left(\frac{v_1}{v_e}\right)^2\right]}. 
\end{eqnarray}
Here $v_1,v_e$ are respectively the initial and terminal velocity, with a default ratio of $v_1/v_e=0.2$. Substituting this, the integral can be analytically solved as
\begin{align}
   \kappa_\lambda & = \dfrac{\tau_{\lambda, \rm DUSTY}}{\dfrac{\dot{M}_{\rm DUSTY}}{4\pi r_1 v_e} \displaystyle \int^{Y}_1\dfrac{1}{y^2\sqrt{1-j/y}}dy} \nonumber \\
    &= \frac{4\pi r_1 v_e\tau_{\lambda, \rm DUSTY}}{ \dot{M}_{\rm DUSTY}}\frac{j}{2}\left(\sqrt{1-\frac{j}{Y}} - \sqrt{1-j}\right)^{\!\!-1},
\end{align}
where $y=r/r_1$ and $j = 1 - (v_1/v_e)^2=0.96$. 
In Figure \ref{fig:opacity&amax}, we show $\kappa_\lambda$ for 9 models of $a_{\rm max}$ from 0.0003 to 0.25 $\mu \text{m}$. We interpolate these curves, linearly in $a_{\rm max}$ and logarithmically in $\lambda$, to obtain $\kappa_\lambda(a_{\rm max})$ at a given time and radius.

\subsection{Light Curve with Dust Correction}
\label{subsec:LC_with_dust}
Given our model for the time-dependent dust formation and its optical depth $\tau_\lambda(t)$, we finally calculate the emerging SED after dust attenuation and reddening. The input SED is given as
\begin{equation}
    \dfrac{dL_{\rm input}}{d \lambda} = \dfrac{dL_{\rm ECSN}}{d \lambda}  + \dfrac{dL_s}{d \lambda}.
\end{equation}
For the ECSN component, we extract the time-dependent SED output from the STELLA calculations in \cite{Sato2026}. For the shock interaction component, we approximate its emission as a thermal SED
\begin{equation}
    \frac{dL_{s}}{d\lambda} = 4\pi r_{bb, Ls}^2\left(\frac{2\pi hc^2}{\lambda^5}\right)\frac{1}{\exp\left[hc/(k_BT_{bb, Ls}\lambda)\right]-1},
\end{equation}
where  $h$ is the Planck's constant, $c$ is speed of light, and $T_{bb, Ls}$ and $r_{bb, Ls}$ are respectively
\begin{align}
    T_{bb, Ls} & \simeq \left( \dfrac{9}{4a_{\rm rad}}\rho_{\rm wind}v_s^2\right)^{1/4} \simeq\left( \dfrac{9L_s}{8\pi r_s^2 v_s a_{\rm rad}}\right)^{1/4}, \\
    r_{bb, Ls} &= \sqrt{\dfrac{L_s}{4\pi \sigma T_{bb, Ls}^4}},
\end{align}
where $a_{\rm rad}$ is the radiation constant. This assumes thermal equilibrium in the post-shock region, with its pressure ($\simeq 3\rho_{\rm wind}v_s^2/4$ at the immediate downstream) being dominated by radiation when it is thermalized. Thermal equilibrium is justified for the slow shock velocities of our case \citep{Svirski12}, and we verified that for our adopted winds radiation pressure indeed dominates over gas pressure by a factor $\gtrsim 10^3$. 

As outlined in the right panel of Figure \ref{fig:modelimage}, the SED of radiation emerging from the dusty wind is modeled as a sum of the fraction of $L_{\rm input}$ that escapes and the fraction absorbed and re-radiated by the dust layer.
At a given time, the SED of the escaping photons is
\begin{eqnarray}
    \left(\frac{dL}{d\lambda}\right)_{\rm esc} = e^{-\tau_\lambda}\frac{dL_{\rm input}}{d\lambda},
\end{eqnarray}
and from energy balance the total luminosity absorbed by dust is
\begin{eqnarray}
    L_{\rm abs} = \int d\lambda~(1-e^{-\tau_\lambda})\frac{dL_{\rm input}}{d\lambda}.
\end{eqnarray}
We model the dust re-emission as a thermal SED, whose temperature is derived as follows. We first search for the dust photosphere $r = R_{d,ph}(\geq r_{\rm in})$ that satisfies
\begin{eqnarray}
\tau_{\lambda_{\rm{pk}}}(r=R_{d,ph})
= \int^{\infty}_{R_{d,ph}} \kappa_{\lambda_{\rm{pk}}}f_{\rm{dust}}\rho_{\rm wind} dr \approx 1,
\end{eqnarray}
where $\lambda_{\rm pk}\approx (2898~{\rm \mu m~K}) ( {L_{\rm abs}}/{16\pi \sigma R_{d,ph}^2})^{-1/4}$ is the peak wavelength for a blackbody spectra. Such solution for $R_{d,ph}$ exists if the dust layer is optically thick to its own emission (peaking in IR), and in this case the radius and temperature of the blackbody emission would respectively be $R_{d,bb}=R_{d, ph}$ and $T_{\rm d, bb}=( {L_{\rm abs}}/{16\pi \sigma R_{d,ph}^2})^{1/4}$, for $Q_{\rm rat} = 1$ as in equation (\ref{eq: rsub_scaling}). 

However, in typical cases we find that the dust layer is optically thin to its own emission. In this case we adopt the blackbody radius to be the innermost dust radius $R_{d,bb}\approx r_{\rm in}$. The dust re-emission $dL_{\rm dust}/d\lambda$ is thus modeled as a thermal SED with temperature equal to the radiative equilibrium temperature $T_{\rm{d,bb}}=(L_{\rm abs}/16\pi \sigma R_{d,bb}^2)^{1/4}$,
\begin{equation}
    \frac{dL_{\rm dust}}{d\lambda} = 16\pi R_{d,bb}^2\left(\frac{2\pi hc^2}{\lambda^5}\right)\frac{1}{\exp(\frac{hc}{k_BT_{\rm d, bb}\lambda})-1}.
\end{equation}

Finally, we take into account the smoothing of the light curve by light travel-time effects (dust echo).
This smoothing would be relevant for the contribution of the input luminosity that either is re-emitted by dust, or undergo scattering before it escapes (see Figure \ref{fig:modelimage}). 

For light rays that freely propagate from an emitting shell with radius $R$, the time lag from a region at angle $\theta~(0 < \theta < \pi)$ with respect to the line towards the observer is expressed as $\Delta t(\theta) = R(1-\cos{\theta})/c$, resulting in $0 < \Delta t<2R/c$. For isotropic emission, light is smoothed uniformly over a time $2R/c$ before reaching us (see e.g. Section 2.1 in \citealt{2015Maeda}). We therefore adopt a box-shape function $\phi(R,t)$ for the distribution of time lag as
\begin{equation}
{\phi(R, t)=}  
\begin{cases}
\dfrac{c}{2R} & (-\dfrac{2R}{c}< t <0) \\
0 & (t < -\dfrac{2R}{c} ~{\rm or}~ 0<t)
\end{cases}
\label{eq:light_travel}
\end{equation}

\begin{figure*}
    \centering
    \includegraphics[width=0.9\textwidth]{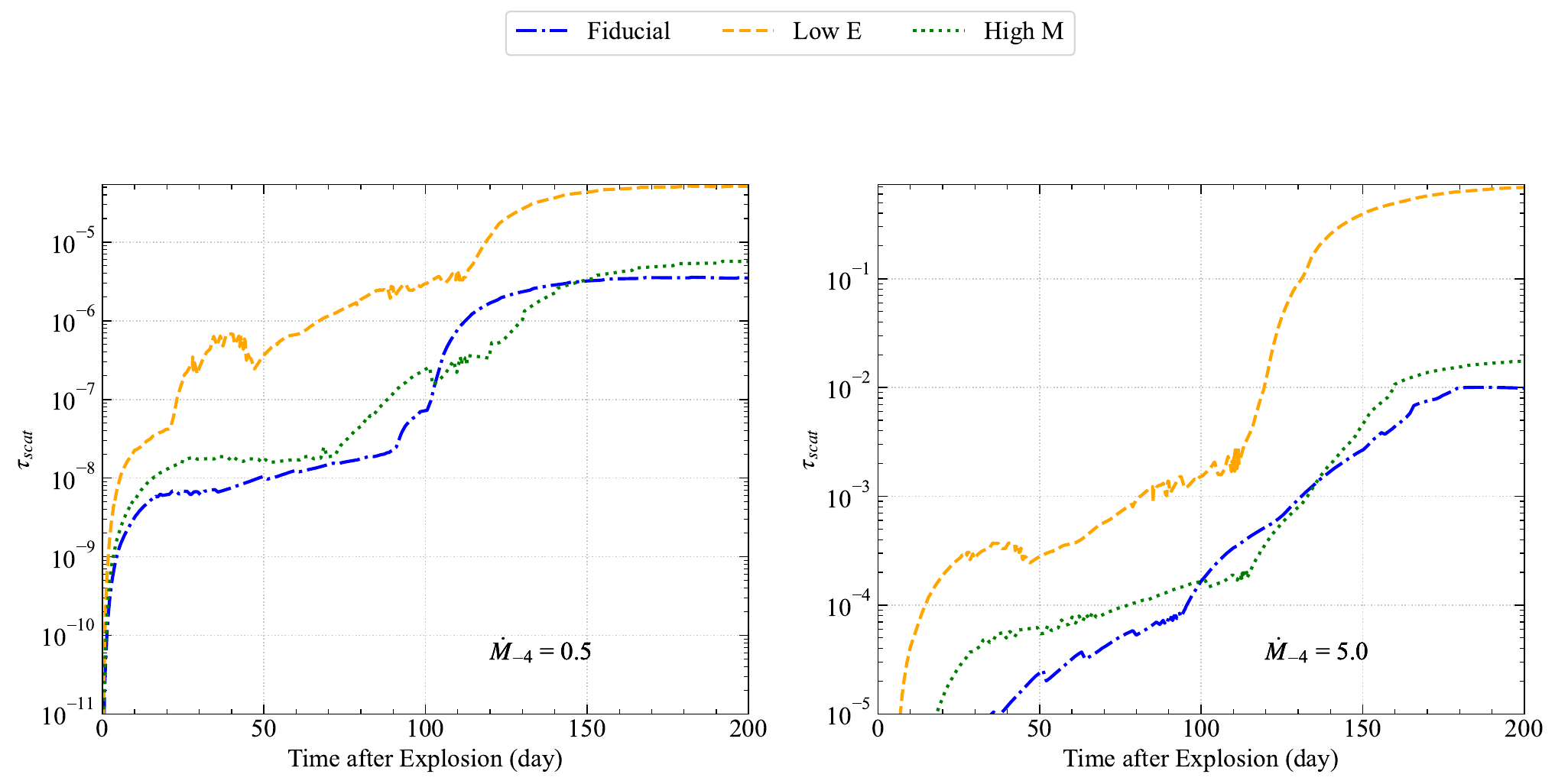}
    \caption{Time evolution of the scattering opacity in the dust-laden layer $\tau_{scat}$ in the V band, for three ECSN models with wind mass-loss rates of $\dot{M}_{-4} = 0.5$ (left) and $5$ (right). While $\tau_{\rm scat}$ rises rapidly after around 100 days when the luminosity and sublimation radius drop, it generally stays below unity for the timescales of our interest.}
    \label{fig:tau_scat}
\end{figure*}
To model the scattered component, we extract the wavelength-dependent albedo ($\alpha_\lambda$) of the dust from DUSTY, with the same interpolations done for $\kappa_{\lambda}(a_{\rm max})$. The scattering opacity $\tau_{\rm scat}$ is then
\begin{equation}
    \tau_{\rm scat} = \int \kappa_{\lambda} \alpha_{\lambda}\rho_{\rm wind }f_{\rm dust} dr.
\end{equation}
Figure \ref{fig:tau_scat} shows the evolution of $\tau_{\rm scat}$ for three ECSN models with $\dot M_{-4} = [0.5, 5]$. Due to the grains having typical sizes much less than the wavelength of optical light, the albedo is small and we find $\tau_{\rm scat}<1$ in the epochs of our interest. 

When $\tau_{\rm scat}<1$, a fraction $\tau_{\rm scat}$ of the escaping emission is smeared over the light-crossing time, whereas a fraction $1-\tau_{\rm scat}$ escapes nearly radially without smearing. When $\tau_{\rm scat}>1$, we expect nearly all the photons to be scattered at least once, and the escaping emission be smeared over the diffusion timescale. While we expect $\tau_{\rm scat}<1$, for completeness we also take into account the case of $\tau_{\rm scat}>1$ and smoothly connect these two regimes by smearing the scattered component over a timescale $\approx {\rm max}(1,\tau_{\rm scat})\times 2r_{\rm in}/c$, as scattering is dominated by the innermost region that can form dust (equations \ref{eq:t_rec}, \ref{eq:tgrow_amin}).

Summarizing the discussions above, we calculate the three components  of luminosity in Figure \ref{fig:modelimage} as,
\begin{align}
    \dfrac{dL_{\rm un-scat}}{d \lambda} &= e^{-\tau_{\lambda}}(1-\tau_{\rm scat})\dfrac{dL_{\rm input}}{d\lambda}  \label{eq:Lunscat}\\
    \dfrac{dL_{\rm scat}}{d \lambda} &= e^{-\tau_{\lambda}}\tau_{\rm scat} \int ^t_{t-\frac{2r_{\rm in}}{c}}dt' \dfrac{dL_{\rm input}(t')}{d\lambda} \dfrac{c}{2r_{\rm in}}\\
    \dfrac{dL_{\rm dust}}{d \lambda} &= \int ^t_{t-\frac{2R_{d, bb}}{c}}dt' 
\dfrac{dL_{\rm dust}(t')}{d\lambda} \dfrac{c}{2R_{d, bb}}
\end{align}
for $\tau_{\rm scat}<1$, and
\begin{eqnarray}
    \dfrac{dL_{\rm un-scat}}{d \lambda} &=& 0\\
    \dfrac{dL_{\rm scat}}{d \lambda} &=& e^{-\tau_{\lambda}}\int ^t_{t-\frac{2\tau_{\rm scat}r_{\rm in}}{c}}dt'\dfrac{dL_{\rm input}(t')}{d\lambda}\dfrac{c}{2\tau_{\rm scat}r_{\rm in}}\label{eq:Lscat}\\
    \dfrac{dL_{\rm dust}}{d \lambda} &=& \int ^t_{t-\frac{2R_{d, bb}}{c}}dt'
    \dfrac{dL_{\rm dust}(t')}{d\lambda}  \dfrac{c}{2R_{d, bb}}
\end{eqnarray}
for $\tau_{\rm scat}\geq 1.$ We finally calculate the optical luminosity of ECSNe after dust attenuation as
\begin{equation}
L_{\rm opt} = \int^{890\rm nm}_{430\rm nm}d\lambda \left( \dfrac{dL_{\rm un-scat}}{d \lambda} + \dfrac{dL_{\rm scat}}{d \lambda}+\dfrac{dL_{\rm dust}}{d \lambda}\right).
\label{eq:Lintegral}
\end{equation}

\begin{figure*}[t]
    \centering
    \includegraphics[width=0.9\textwidth]{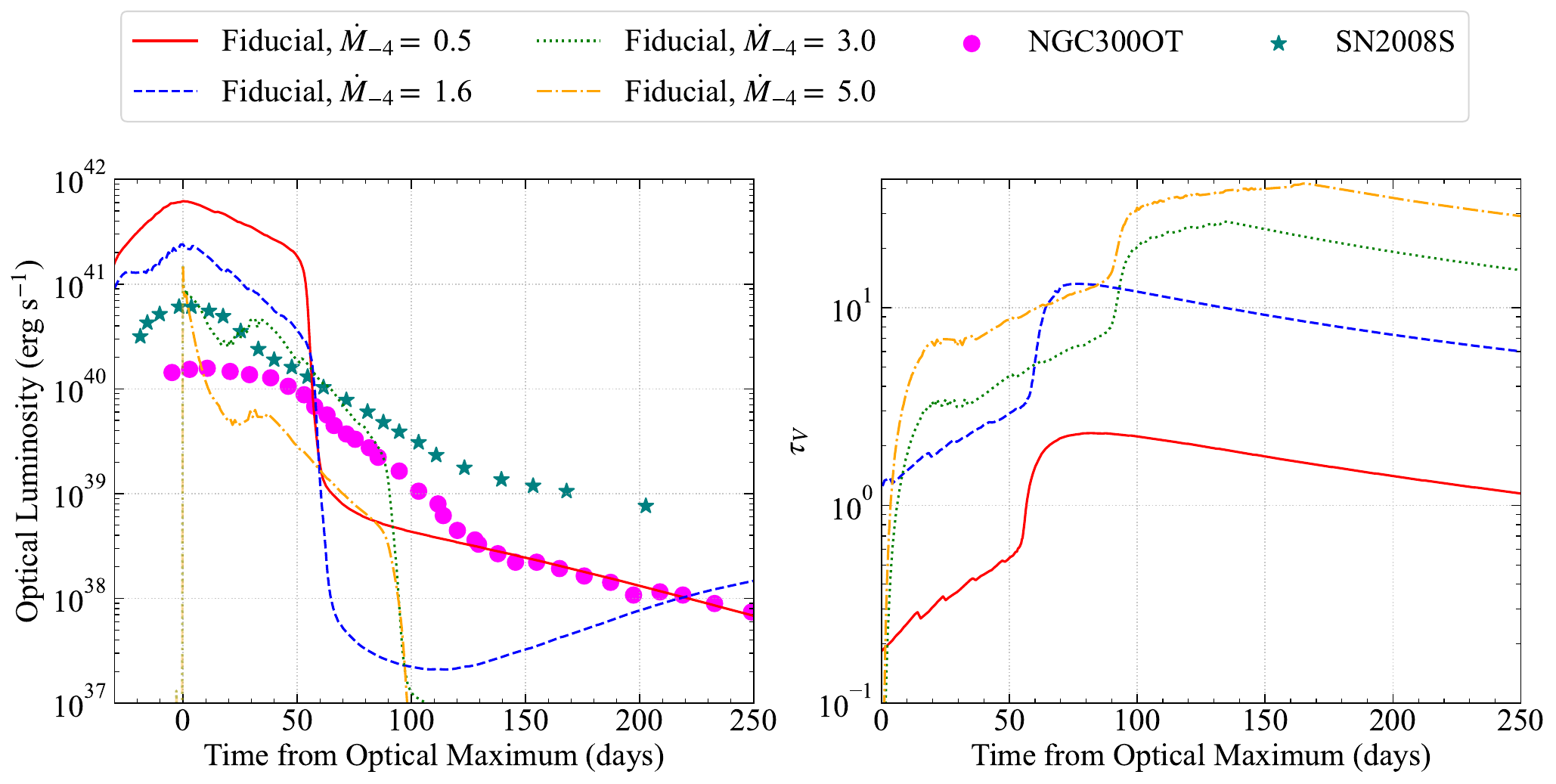}
    \caption{Left: Light curves after dust attenuation for the Fiducial ECSN model, for a range of $\dot{M}_{-4}=0.5$--$5$. Data points show the optical light curves of two representative ILRTs with late-time photometric data, SN 2008S and NGC 300-OT \citep[taken from][]{Cai2021}. Right: Time evolution of optical depth in the V band, for the same values of $\dot{M}_{-4}$. For the time axis, we used the time from the optical maximum of each model with a different mass-loss rate. For models with higher $\dot{M}_{-4}=3,5$, dust reforms within days reaching $\tau_V >1$. These models reach an optical maximum shortly after the shock breakout and are soon attenuated by the rapidly reforming dust, resulting in a light curve starting with a peak near shock breakout. The luminosity before optical maximum for these cases represents the luminosity before shock breakout.}
    \label{fig:LC Comparison Ls analytical}
\end{figure*}
\section{Results}
\label{sec:result}
\subsection{Comparison to ILRT light curves}
\label{subsec:LCcomparison_Ls_Analytical}
Figure \ref{fig:LC Comparison Ls analytical} shows the optical light curves for our fiducial ECSN model, with four wind mass-loss rates $\dot{M}_{-4} = [0.5, 1.6, 3.0, 5.0]$. In the plateau phase, for $\dot{M}_{-4} = 0.5, 1.6$ the effect of dust is relatively small and the optical light curve is similar to the one of ECSN with peak during the plateau phase. On the other hand, when $\dot{M}_{-4} = 3, 5$, the rapid dust growth results in a quick rise to $\tau_V\gg 1$, leading to strong dust extinction in their plateau phase. This results in having peak shortly after the shock breakout, followed by a drop in the optical light curve due to dust attenuation.

In the transition from plateau to tail, $L_{\rm input}$ drops drastically as shown in Figure \ref{fig:analytical Ls}. Then the sublimation radius decreases by at least a factor of a few, accelerating dust formation and growth and leading to a sharp rise in $\tau_V$. 
This occurs $\sim 50$ days after optical maximum for $\dot{M}_{-4} = 0.5,1.6$, and $\sim 90$ days after optical maximum for $\dot{M}_{-4} = 3.0,5.0$. The combination of the intrinsic sharp drop in $L_{\rm input}$ and the sharply rising $\tau_V$ results in the optical luminosity plummeting by orders of magnitude. 

After that, as the shock radius increases, the reformed dust is destroyed by the shock, and $\tau_V$ then slowly drops with time. This leads to an increase in the light curve for $\dot{M}_{-4} = 1.6$, with the lag between the peak of $\tau_V$ and the luminosity minimum set by light-travel effects. Such late rise is not seen for $\dot{M}_{-4} = 0.5$, since the decline in $\tau_V$ is not significant and does not overcome the decline in the input luminosity.

Comparing our result with the optical light curves of two representative ILRTs, SN 2008S and NGC 300-OT (left panel of Figure \ref{fig:LC Comparison Ls analytical}), we find large discrepancies in their light curve evolutions. For small mass loss rates ($\dot{M}_{-4}\lesssim 1$, such as the solid line), the luminosity around optical peak ($\sim 6\times10^{41}\rm erg~ s^{-1}$) is still much larger than ILRTs ($\sim 7\times10^{40}\rm erg~ s^{-1}$ for SN2008S and $\sim 2\times10^{40}\rm erg ~s^{-1}$ for NGC300OT). More generally, the luminosity drops too steeply during the transition from the plateau to the tail around 60 days after optical maximum, and fails to reproduce the smooth evolution of ILRTs. The luminosity drop becomes more drastic for larger mass loss rates, making them essentially invisible in the optical after $\approx 100$ days for $\dot{M}_{-4}\gtrsim 3$ (dotted line and dash-dotted line in the left figure of Figure \ref{fig:LC Comparison Ls analytical}). 

We find that the root cause of this discrepancy in the optical light curves is the sharp luminosity drop in $L_{\rm input}$ (Figure \ref {fig:analytical Ls}), which rapidly increases $\tau_V$ and further drops the optical luminosity. We may consider two scenarios to suppress this sharp transition in $L_{\rm input}$ and/or the dust optical depth -- a further increase in shock luminosity to smoothen the drop, or an additional contribution in optical depth from dust that survived the shock breakout. Hereafter we discuss each of these scenarios.

\subsection{Increase in shock luminosity}
\label{subsec: increase shock L}

\begin{figure*}[t]
    \centering
    \includegraphics[width=0.9\textwidth]{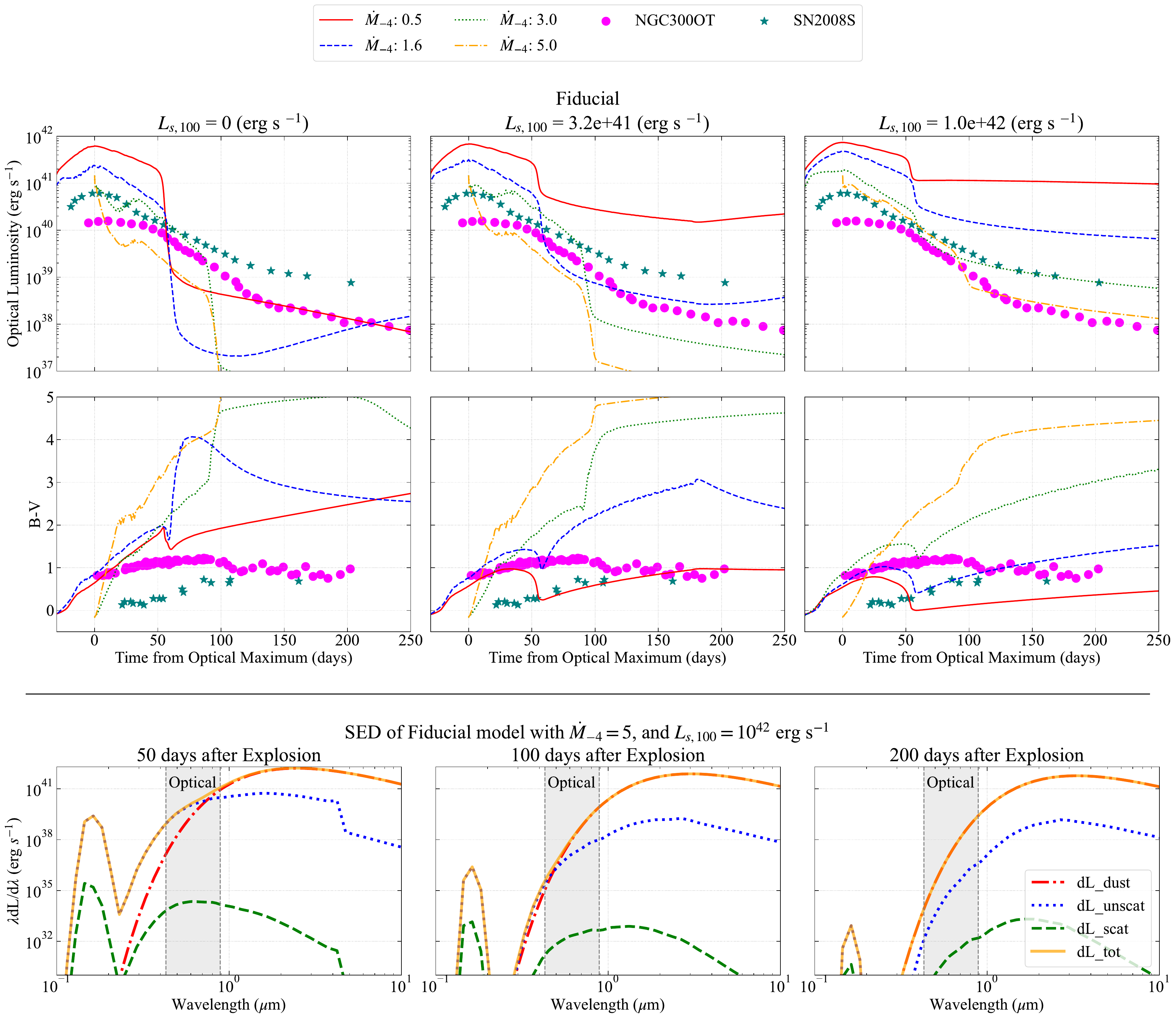}
    \caption{Luminosity (top) and color $B-V$ (middle) evolutions, for the Fiducial ECSN model with additional shock luminosity input. The added shock luminosity parameter $L_{\rm s, 100}$ (equation \ref{eq:Ls_rad_additional}) is varied as $0$ (left; reference model same as Figure \ref{fig:LC Comparison Ls analytical}), $3.2\times10^{41}~\rm erg~s^{-1}$ (middle), and $10^{42}~\rm erg~s^{-1}$ (right). Lines correspond to different values of $\dot{M}_{-4}$, and data points are the observed luminosity and color of NGC300-OT and SN 2008S \citep{Botticella2009, Humphreys2011,Cai2021}. Bottom panels show the SED evolution of the model with $(\dot{M}_{-4},L_{s, 100}) = (5,10^{42}~\rm erg~s^{-1})$ at [50, 100, 200] days after explosion. While the wavelength-integrated optical light curve roughly agrees with the observed ILRTs, the increasing contribution from dust re-emission makes the SED in the optical (gray shared region) much redder than observed.}
    \label{fig:LC_ColorFoducial}
\end{figure*}

As the ECSN plateau gives a contribution to the input luminosity of nearly $10^{42} ~\rm erg~s^{-1}$, a similar contribution from CSM interaction could suppress the sharp drop in the input luminosity. While $\dot{M}_{-4}$ around unity expected for super-AGB winds cannot realize this (Figure \ref{fig:analytical Ls}), this may be possible if the progenitor had a much higher mass-loss rate just before core-collapse\footnote{This may also be supported from the prominent narrow Balmer lines commonly seen in ILRTs \citep{Cai2021,Valerin2025_spectra}, although to our knowledge quantitative estimates of the mass loss rate from spectral modeling do not exist.}. 
Here we consider a modification of the density profile, which now constitutes of a stellar wind and a confined CSM of an enhanced mass loss rate $\dot{M}_{\rm dense}$ within radius $r_{\rm dense}$,
\begin{equation*}
    \rho_{\rm{wind}}(r)=
    \begin{cases}
        \dot{M}_{\rm{dense}}/{4\pi r^2 v_w} & (r < r_{\rm{dense}}) \\
        \dot{M}/{4\pi r^2 v_w} & (r \geq r_{\rm{dense}}).
    \end{cases}
\end{equation*}
From Section \ref{subsec: Input Luminosity} one can estimate the required $\dot{M}_{\rm dense}$ to power a shock luminosity of $L_{s,100}$ at 100 day from explosion (around the end of the plateau) to be 
\begin{align}
\dot{M}_{\rm dense} \simeq 9.7\times 10^{-3}~M_{\odot}~{\rm yr}^{-1}\left(\frac{L_{s, 100}}{10^{42} ~\rm erg~ s^{-1}}\right)^{\!\!10/7} \nonumber \\
\left(\dfrac{E_{\rm ej}}{10^{50}~\rm erg}\right)^{\!\!-27/14}\left(\dfrac{M_{\rm ej}}{5~M_{\odot}}\right)^{\!3/2}.
\label{eq:M_dense}
\end{align}

Before presenting detailed calculations of the light curves, we comment on the range of $r_{\rm dense}$ where this scenario could plausibly work in the context of ILRTs. First, in order to suppress the sharp light curve drop, the enhanced shock power needs to be maintained over the whole observed duration. From equation (\ref{eq:r_sh_scaling}) a constraint of $r_{\rm dense} > r_{\rm s}(t)$ over $t\lesssim 1$ yr translates to a lower limit of $r_{\rm dense} \gtrsim$ (a few) $\times 10^{15}~\rm cm$, which agrees with more detailed numerical calculations of \cite{Sato2024}.
Second, we can see in Figure \ref{fig:LC Comparison Ls analytical} that simply extending the dense CSM beyond the sublimation radius would highly obscure the optical emission. This hence requires $r_{\rm dense} < r_{\rm sub}$, so that the dense CSM will not form dust and contribute to the dust optical depth. From equation (\ref{eq: rsub_scaling}), we expect an upper limit of $r_{\rm dense} \lesssim 10^{16}~\rm cm$. While the two limits do not contradict, this requires $r_{\rm dense}$ to be narrowly tuned, such that this enhanced mass loss must have occurred in the last few 100 years before the SN for $v_w=10\ {\rm km~s^{-1}}$. 

Under such a restricted range of $r_{\rm dense}$, the dust optical depth is still governed by the lower-density wind, and a dense confined CSM would simply enhance the shock luminosity and $L_{\rm input}$. As $L_{\rm s,rad}=L_s \propto t^{-0.3}$ for the high-density CSM (see Section \ref{subsec: Input Luminosity} and equation \ref{eq:L_shock}), we capture this effect by introducing an additional shock luminosity in our model\footnote{We note that an enhanced $\dot{M}_{\rm dense}$ also leads to prolonged shock breakout emission due to photon diffusion, that would modify the earliest light curve over the diffusion timescale, $\approx \kappa\dot{M}_{\rm dense}/4\pi v_w c\sim 6\ {\rm day}~(\kappa/0.3\ {\rm cm^2\ g^{-1}})(\dot{M}_{\rm dense}/10^{-2}M_\odot\ {\rm yr}^{-1}$). While this is not captured in our model for $L_{s, \rm rad}$, we expect this to not affect our conclusions that mainly concern the evolution around the end of the plateau ($\gtrsim 100$ days from explosion).}
\begin{equation}
L_{s,\rm rad}(t) = L_{s,100} \times \Big(\frac{t}{100\ {\rm days}} \Big)^{-0.3}.
\label{eq:Ls_rad_additional}
\end{equation}
For a given $L_{s, 100}$, we use the corresponding $\dot{M}_{\rm dense}$ obtained from equation (\ref{eq:M_dense}) to calculate the shock propagation and SED from shock interaction. Then we adopt the lower $\dot{M}$ for calculating the dust growth and extinction/re-emission.

In the top panel of Figure \ref{fig:LC_ColorFoducial}, we compare the resulting light curves with those of SN 2008S and NGC 300-OT. We now have two free parameters in our models ($\dot{M}_{-4}$, $L_{s, 100}$), which we vary as $\dot{M}_{-4} = [0.5, 1.6, 3.0, 5.0]$ and $L_{s,100} = [0,3.2,10]\times10^{41} ~\rm erg~s^{-1}$. Here $L_{s, 100} = 0$ corresponds to a reference model with no additional shock power from the analytically derived $L_{s, \rm rad}$ in Section \ref{subsec: Input Luminosity}, i.e. the same as Section \ref{subsec:LCcomparison_Ls_Analytical} and Figure \ref{fig:LC Comparison Ls analytical}. Here we focus on the case of the Fiducial ECSN model in Table \ref{table:model_info}, and show plots for other ECSN models in the Appendix (which result in qualitatively similar conclusions).

Comparing different values of $L_{s, 100}$, adding additional shock luminosity of $L_{s,100}\sim 10^{42}$ erg s$^{-1}$ removes the drastic drop in luminosity at the end of the plateau, allowing the light curves to have a tail similar to the observed ILRTs. As the plateau luminosity is nearly unaltered, we still find that a wind mass loss rate of $\dot{M}_{-4}\approx 5$ can decrease the early-time luminosity down to those of ILRTs. Then a model of $L_{s,100}=10^{42}$ erg s$^{-1}$ provides the best match in the late-time optical luminosity.

However, comparing the wavelength-integrated optical light curve is not sufficient. Since the dust that attenuates the plateau luminosity also converts the input luminosity into redder wavelengths via dust re-emission, a luminosity match does not guarantee a color match. Therefore, we also compare the optical color $B-V$ between our model\footnote{The $B$ and $V$-band magnitudes are calculated by applying to our total SED the Bessel-B and Bessel-V band filter profiles (from \url{http://svo2.cab.inta-csic.es/svo/theory/fps}; \citealt{Rodrigo12,Rodrigo20,Rodrigo24}), and converting to the Vega system.} and ILRTs, in the middle row panels of Figure \ref{fig:LC_ColorFoducial}.

We find that models with large $\dot{M}_{-4}$ in order to match the plateau luminosity results in extremely red colors (large $B-V$) after the plateau. For example, all models with $\dot{M}_{-4}= 5$ and varied different shock luminosity parameters result in $B-V > 4$ in the tail phase, much redder compared to ILRTs at similar phases. 

To understand this strong reddening, we discuss how the three luminosity components in our model evolve over time. The bottom row of Figure \ref{fig:LC_ColorFoducial} shows the SEDs for the model with $(\dot{M}_{-4}, L_{s,100})= (5,10^{42} ~\rm erg~s^{-1}$) across three main phases: the early part of the plateau, the transition from the plateau to the tail, and the tail phase. For each epoch, we show the SEDs for each of the three components as well as the total SED. We find that as time passes, the contribution of $L_{\rm scat}$ and $L_{\rm un-scat}$ (escaping photons) decreases due to dust reformation, and dust re-emission becomes more dominant in the optical SED at later times. As dust re-emission is a thermal SED peaking in the IR, we conclude that dust re-emission sets this large $B-V$. As the dust optical depth is fairly sensitive to the mass loss rates, this reddening and re-emission becomes stronger for larger mass loss rates. This qualitative trend is also seen for the other two ECSN models. (see Figures \ref{fig:LCColor_lowE} and \ref{fig:LCColor_highM}). We thus conclude that adding an enhanced shock luminosity component cannot reproduce both the light curve and color evolution of ILRTs.

\subsection{Contribution from surviving dust}
\label{sec:surviving_dust}
\begin{figure*}[t]
    \centering
    \includegraphics[width=0.9\textwidth]{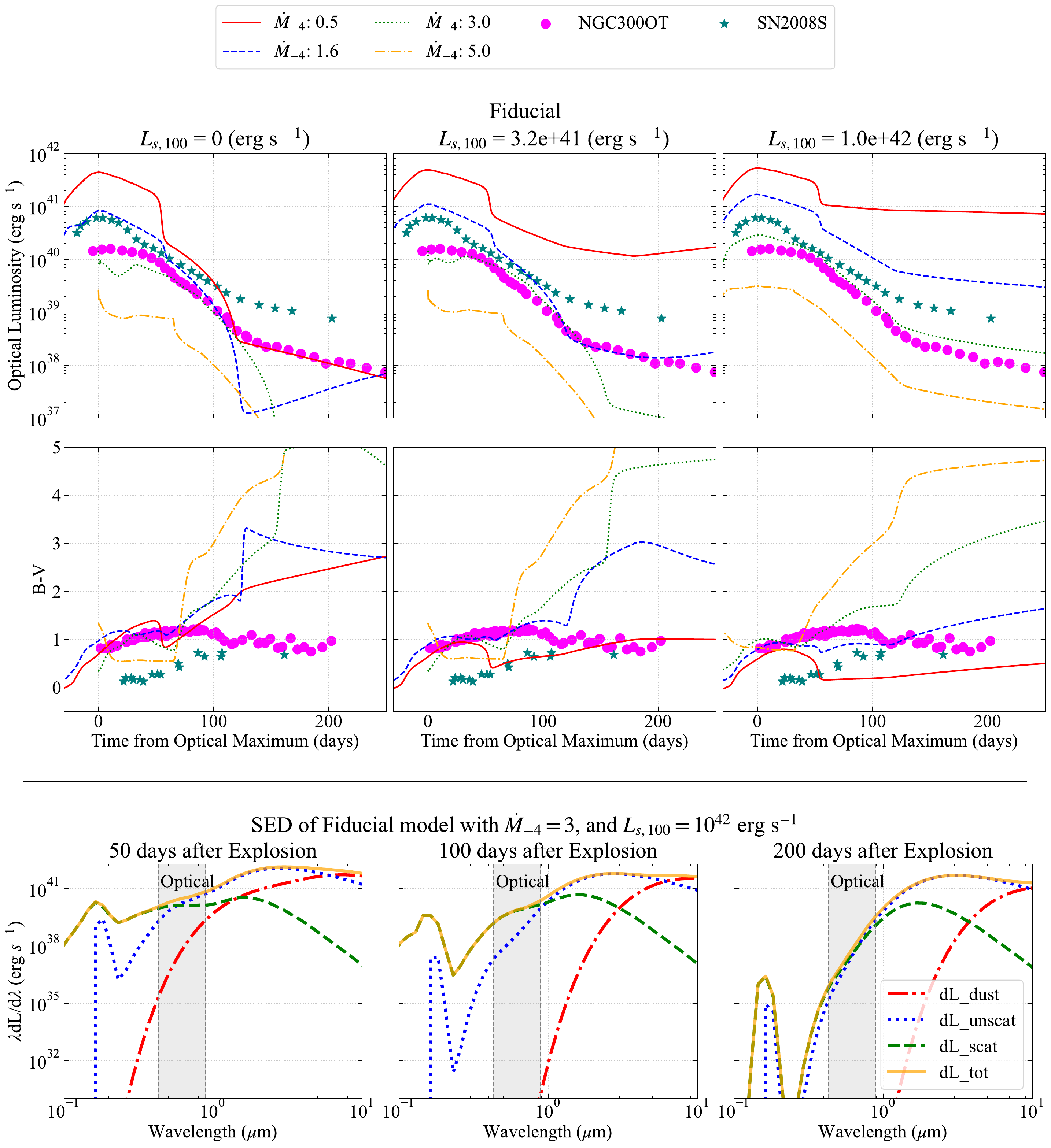}
    \caption{Same as Figure \ref{fig:LC_ColorFoducial} (Fiducial model), but including the additional attenuation by the dust surviving shock breakout. Here the four components in the SED (bottom panel) correspond to those from reprocessing by the outer surviving dust.}
    \label{fig:SurvivingDust_Fiducial}
\end{figure*}

Our calculations so far have not included the possible effects of outermost dust that can survive from the shock breakout emission.
The shock breakout from ECSNe (with luminosity $L_{\rm sb}\sim 10^{44}$ erg s$^{-1}$; Figure \ref{fig:LCmodel}) can destroy dust out to a radius
\begin{equation}
r_{\rm dest} \simeq 8\times 10^{16}~{\rm cm} \left( \dfrac{L_{\rm sb}}{10^{44}~\rm erg~ s^{-1}}\right)^{1/2}. \label{eq: destruction radius}
\end{equation}
Given the theoretically expected duration of the super-wind phase of $\sim 10^4$--$10^5$ years \citep{Poelarends08,2017Doherty} dust may exist well outside this radius, though whether the wind will have a similar mass loss rate is less certain. If the dusty wind at $r\gg r_{\rm dest}$ extends with the same $\dot{M}$, its optical depth can be roughly estimated as
\begin{eqnarray}
    \tau_{\rm V}&\approx& \frac{\kappa_V\dot{M}}{4\pi v_w r_{\rm dest}} \nonumber \\
    &\sim& 0.6~\dot{M}_{-4}\left(\frac{\kappa_V}{100\ {\rm cm^2\ g^{-1}}}\right)\left(\frac{r_{\rm dest}}{8\times 10^{16}\ {\rm cm}}\right)^{\!\!-1},
    \label{eq: tau-survive}
\end{eqnarray}
where $\kappa_V$ is the V-band opacity. Comparing this with Figure \ref{fig:LC Comparison Ls analytical}, this can be a significant early contribution to the dust extinction. Such addition can lessen the required $\dot{M}$ to match the plateau with ILRTs, and may hence alleviate the issue of the aforementioned strong reddening at late times.

To investigate the effect of pre-existing dust, we make the following updates to our light curve model. We assume that this dust has maximally grown to a grain size distribution of a classic MRN distribution with $a_{\rm max} = 250~\rm nm$. The wavelength-dependent optical depth of this surviving dust is set to be time-independent after the explosion, with 
\begin{eqnarray}
 \tau_{\rm survive}(\lambda) &=& \int^{\infty}_{r_{\rm dest}}\kappa_{\lambda}\rho_{\rm wind}dr, \\
 \tau_{\rm scat,survive}(\lambda) &=& \int^{\infty}_{r_{\rm dest}}\kappa_{\lambda}\alpha_\lambda \rho_{\rm wind}dr,
\end{eqnarray}
where $r_{\rm dest}$ is calculated from equation (\ref{eq: destruction radius}) with $L_{\rm sb}$ given by the peak bolometric luminosity at shock breakout (see Table \ref{table:model_info}). To embed this existing dust in our model, we adopt a dual dust shell structure such that the light is attenuated by both the inner reformed dust and the outer surviving dust. To model the latter attenuation, we adopt the output SED ${dL_{\rm tot}}/{d\lambda}$ calculated in Section \ref{subsec:LC_with_dust} as the new ``input luminosity" ${dL_{\rm input}}/{d\lambda}$. 

With this new input, we repeat the calculation of three components, $L_{\rm un-scat}, L_{\rm scat}$, and $L_{\rm dust}$, using $\tau_{\rm survive}, \tau_{\rm scat, survive}$, and $r_{\rm dest}$ which are all constants of time. 
The smearing of light curve by the light travel effects is also considered for this surviving dust. The scattered components are smoothed by $\max(1, \tau_{\rm scat, survive})\times2r_{\rm dest}/c$, whereas dust re-emission components are smoothed by $2r_{\rm dest}/c$.

Figure \ref{fig:SurvivingDust_Fiducial} is the same as Figure \ref{fig:LC_ColorFoducial}, but including the effect of surviving dust (see Figure \ref{fig:SurvivingDust_lowE} and \ref{fig:SurvivingDust_highM} for the other two ECSN models). Since the optical depth of surviving dust is constant over time and scales with the mass-loss rate (equation \ref{eq: tau-survive}), the light curves experience additional attenuation over the entire phase, with stronger effects for higher $\dot{M}_{-4}$. This allows us to suppress the plateau luminosity with a lower $\dot{M}_{-4}$.

In addition, since the scattered component is smoothed by $\sim 2r_{\rm dest}/c\sim60~\rm days$, this component is dominant even after the plateau phase ends (See Figure~\ref{fig:SurvivingDust_Fiducial}, bottom row). This results in an ILRT-like color ($B-V$) evolution until $\sim$ 100 days. However in the late tail, we still systematically find strong reddening of $B-V$ $\gtrsim 3$ for the cases of $(\dot{M}_{-4}, L_{s, 100})$ that broadly reproduce the total optical luminosity. This occurs because the optical luminosity at late times is still dominated by the dust re-emission from the inner reformed dust, which is further reddened by the outer surviving dust to even larger $B-V$ that is inconsistent with ILRTs\footnote{We note an enhanced $\dot{M}_{\rm dense}$ in the middle and right panels could alter the shock breakout emission that is not captured in this model. Examining models of \cite{Sato2026} with and without CSM, we find that including a dense CSM prolongs the breakout and reduces the peak luminosity, which results in a lower radius of the surviving dust with larger optical depth by a factor of a few. However we expect this to not alter our qualitative picture of requiring strong reddening in order to reproduce the optical luminosity at late times.}.

\section{Summary and Discussion}
\label{sec:summary and discussion}
We have constructed a light curve model combining theoretical ECSN light curves with dust extinction expected for super-AGB star winds, to consider whether ECSNe can explain the light curves of ILRTs. Before discussing the implications of our work, we summarize our main findings below.
\begin{itemize}
\item To suppress the bright plateaus in ECSNe down to the luminosities of ILRTs, the dust optical depth in the optical needs to be at least a few in the plateau phase, requiring $\dot{M}_{-4}$ well above unity (see Figure \ref{fig:LC Comparison Ls analytical}). Furthermore, we observe a drop in luminosity by at least factor of 5 right after the plateau phase (see Figure \ref{fig:analytical Ls}). This drop would be further emphasized with the effect of dust, since the dust inner radius shifts inward and leads to a rapid increase in the optical depth. A much larger late-time shock luminosity $L_{\rm s, 100} \sim10^{42} ~\rm erg~s^{-1}$, potentially due to confined CSM with an enhanced mass loss just before core-collapse, is required to suppress the drastic luminosity drop (e.g., top row of Figure \ref{fig:LC_ColorFoducial}).
\item Even by including such large $L_{\rm s, 100}$, every model with a high mass-loss rate ($\dot{M}_{-4}\approx 3.0$--$5.0$) to fit the optical light curve of ILRTs fails to reproduce their color evolution. The calculated $B-V \gtrsim 3$ in the tail phase is much redder compared to $B-V\lesssim 1.2$ in the two events SN 2008S and NGC 300-OT, as well as $B-V\lesssim$ 1.6 in a larger sample of ILRTs (Figure 6 of \citealt{Valerin2025}). This is because most of the optical luminosity at this phase is from absorption and re-emission by dust at temperatures ($T_d \lesssim 1500 ~\rm K$) that peak in the IR (see middle and bottom rows of Figure \ref{fig:LC_ColorFoducial}).
\item Including the effect of the surviving dust after shock breakout reduces the required mass-loss rate (see Figure \ref{fig:SurvivingDust_Fiducial}). In addition, scattering by this outer dust smooths the light curve by $\approx 60~\rm days$, making the light curve more smooth even after the end of the plateau phase. This results in ILRT-like color $B-V$ evolution until $\sim$ 100 days after optical peak; however strong reddening is still seen ($B-V \gtrsim 3$ around 200 days.) This reddening occurs since the dust re-emission would still make the reprocessed SED peak in the IR at late times, and the attenuation by surviving dust also shifts the peak of the SED to even redder wavelengths.

\end{itemize}

These inconsistencies pose a challenge to simultaneously fit both the luminosity and color evolutions under standard ECSN light curves with dust effects, which is the case for all three ECSNe models by \cite{Sato2026} adopted in this work. This suggests that in order to reproduce ILRTs, the transient light curves have to be intrinsically dim without strong effects of dust, likely requiring a much weaker explosion than canonical ECSN predictions ($\sim 10^{50}$ erg). From extrapolation of the scaling relations in Type II-P SNe \citep{Popov93}, \cite{Tominaga13} argued that such weak explosions of $E_{\rm ej}\sim$ (a few) $\times 10^{48}$ erg are required to reproduce the luminosity of SN 2008S near peak. Given that SN 2008S is around the middle in the luminosity range of observed ILRTs \citep{Valerin2025}, a separate population of weak explosions with energies of $10^{48}$--$10^{49}$ erg may be required for ILRTs. 

Recently, it has been suggested that electron capture in an ONeMg core could end up with a thermonuclear explosion leaving a massive white dwarf remnant, instead of the conventional wisdom of a neutrino-driven explosion leaving a neutron star remnant \citep[e.g.,][]{Isern91,Jones16,Jones19,Holas2026}. 
In such scenarios of ``thermonuclear ECSNe (tECSNe)", the explosions are qualitatively similar to the ``pure deflagration" scenarios in Type Iax SNe \citep{Fink14,Kromer15,Lach22}, whose predicted energies overlap with the $10^{48}$--$10^{49}$ erg suggested in this work. Detailed light curve predictions under this scenario and comparisons with ILRTs could be an interesting future work. A recent work of \cite{Holas26b} studied the radiative signatures of tECSNe with explosion energies of order $\sim 3\times 10^{49}$ erg, finding strong Ne II line emission at 12.8~$\mu$m at late times. Such a prediction may be testable by late-time IR spectroscopic follow-up of nearby ILRTs.

We may also consider other progenitor channels that may explode with energies of $\sim 10^{48}$--$10^{49}$ erg, such as stellar outbursts, accretion-induced collapse (AIC) of white dwarfs, or failed SNe forming black holes (BHs). While stellar outbursts can have a range of explosion energies with diverse light curves \citep[e.g.,][]{Dessart10}, it may have difficulty explaining the late-time IR observations of some ILRTs supporting a terminal explosion \citep{Adams2016, Valerin2025}. 

Recent simulations for AIC find explosion energies as low as $10^{46}$--$10^{48}$ erg \citep{Mori26}, and the hydrogen-rich CSM inferred from spectra could be related to the ongoing mass transfer that triggers the AIC. While ILRTs are preferentially observed in star-forming galaxies \citep{Cai2021,Karambelkar23}, AIC can occur in an extremely broad range of delay times up to $\gtrsim$ Gyr \citep[e.g.,][]{Ruiter19}. This may be tested with larger samples of ILRTs expected from Rubin Observatory.

Finally, failed SNe have been proposed to accompany weak explosions, due to the reduction in central gravitational mass due to the neutrino emission \citep{Nadyozhin80,Lovegrove13,Fernandez18} or due to random infall of the convective envelope with large (specific) angular momentum \citep{Antoni23,Antoni26,De26}. Their estimated explosion energies span a range of $\sim 10^{47}$--$10^{49}$ erg, which may align with ILRTs. \citet{Tsuna2020} suggests that failed SNe occurring in dense CSM can explain some ILRTs like AT 2017be. However the study was limited to BH formation from blue supergiants and Wolf-Rayet stars, and broader progenitor channels should be considered for future studies in the context of ILRTs. Additionally, late-time X-ray follow-up of ILRTs may be helpful to test the presence of an accreting BH remnant.

We conclude by discussing possible uncertainties in the model. In this work, we assumed that the dust is 100$\%$ amorphous carbon, with a grain size distribution represented by the MRN distribution \citep{1977Mathis}. While variations in dust chemistry and size distribution can affect the extinction and resulting light curves, we stress that the main reason for the discrepancy between ECSNe and ILRTs is either the sharp drop in the ECSN light curves or strong reddening due to dust re-emission. These effects are largely insensitive to the details of dust. 

Another simplification we have made is that the dust is spherically symmetric, as expected for a stellar wind. If the dust extinction has strong asymmetries, for example the optical depth being larger at the equator than the poles, we may expect varying degrees of extinction due to viewing angle. However such scenarios are not straightforward to strongly obscure the optical emission from ECSNe \citep{Kochanek2024}. We also favor an intrinsically dim explosion to suppress the reddening from dust re-emission, that is expected to be nearly isotropic. Thus we conclude asymmetry will not fully resolve the discrepancy in our ECSN models with ILRTs.
\section*{}
We thank members of the Transient Theory group at Harvard University for support and discussions. Y.T. expresses gratitude to the Institute for Theory and Computation and Grinnell College for their support. 
D.T. is supported by Harvard University through the Institute for Theory and Computation Fellowship.
D.H. is supported by STScI grants HST-GO-17770.002, JWST-GO-12468.001, and JWST-GO-09964.001.
M.S. is supported by the Grant-in-Aid for the JSPS Fellows (grant No. 26KJ0086).
Y.T. acknowledges the use of Claude Fable 5 (Anthropic) for English refinement and the organization and refinement of the code used to generate the figures. All the AI-assisted changes were independently reviewed and examined by the authors, who take full responsibilities for the entire content of their work.
\begin{appendix}
\restartappendixnumbering
\section{Additional Figures}
Here we show light curves from the Low E and High M models in Table \ref{table:model_info}, where we find similar conclusions on its difficulty in reproducing both the luminosity and color evolutions of ILRTs.
 \begin{figure*}[t]
    \centering
    \includegraphics[width=0.9\textwidth]{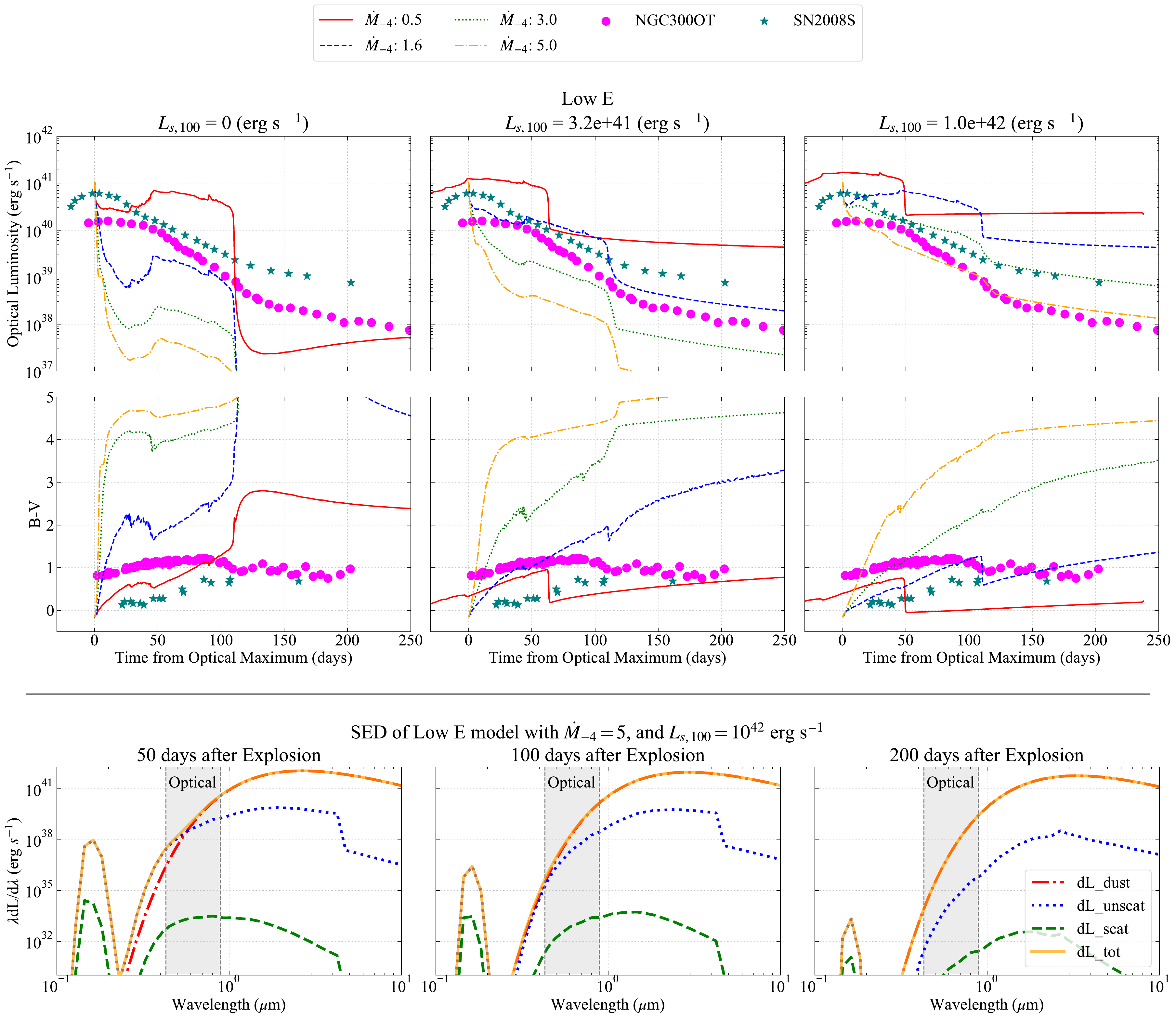}
    \caption{Same as Figure \ref{fig:LC_ColorFoducial} that explored the effect of additional shock luminosity input, but for the Low E model.}
    \label{fig:LCColor_lowE}
\end{figure*}
\begin{figure*}[t]
    \centering
    \includegraphics[width=0.9\textwidth]{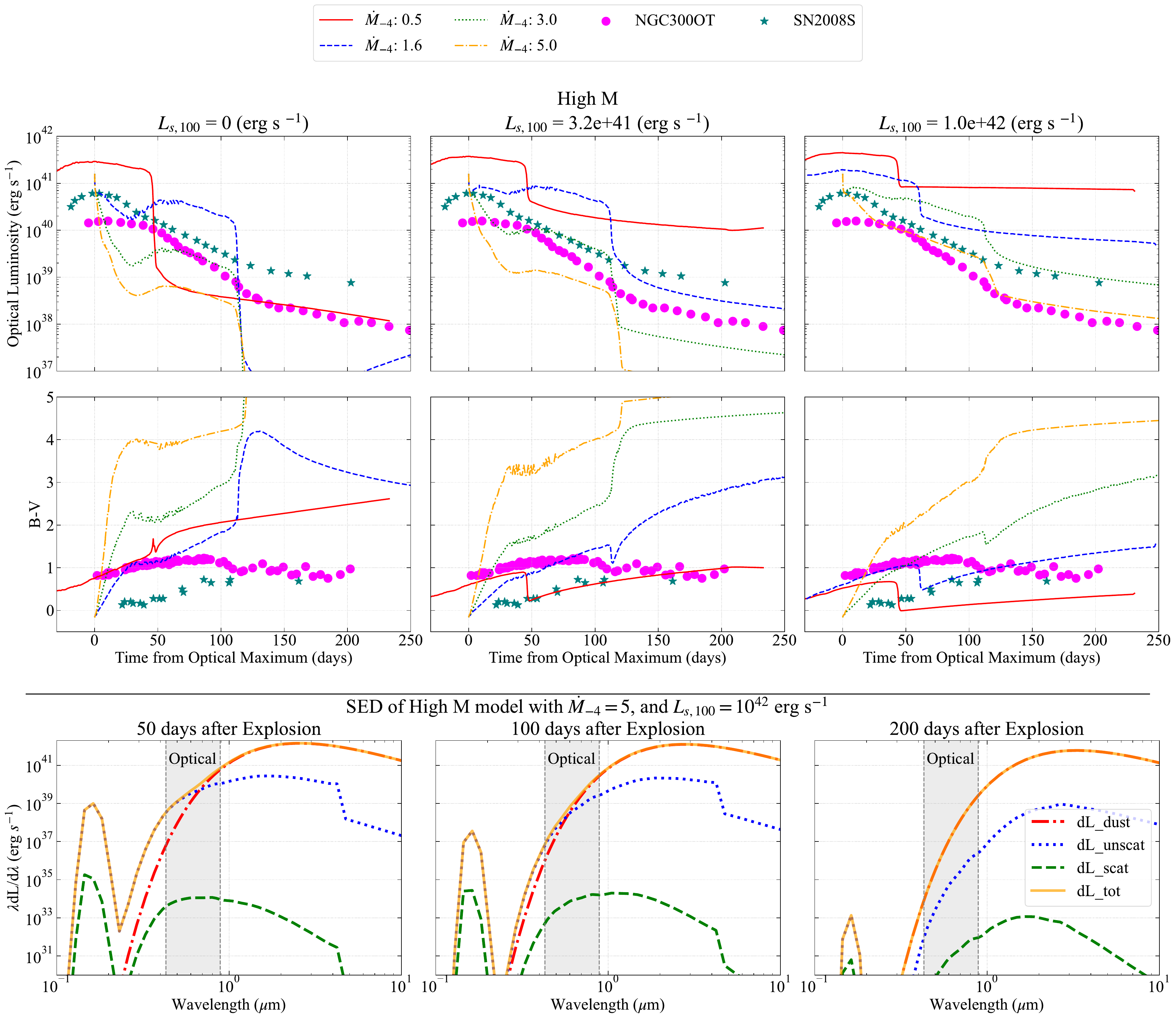}
    \caption{Same as Figure \ref{fig:LC_ColorFoducial} that explored the effect of additional shock luminosity input, but for the High M model.}
    \label{fig:LCColor_highM}
\end{figure*}
\begin{figure*}[t]
    \centering
    \includegraphics[width=0.9\textwidth]{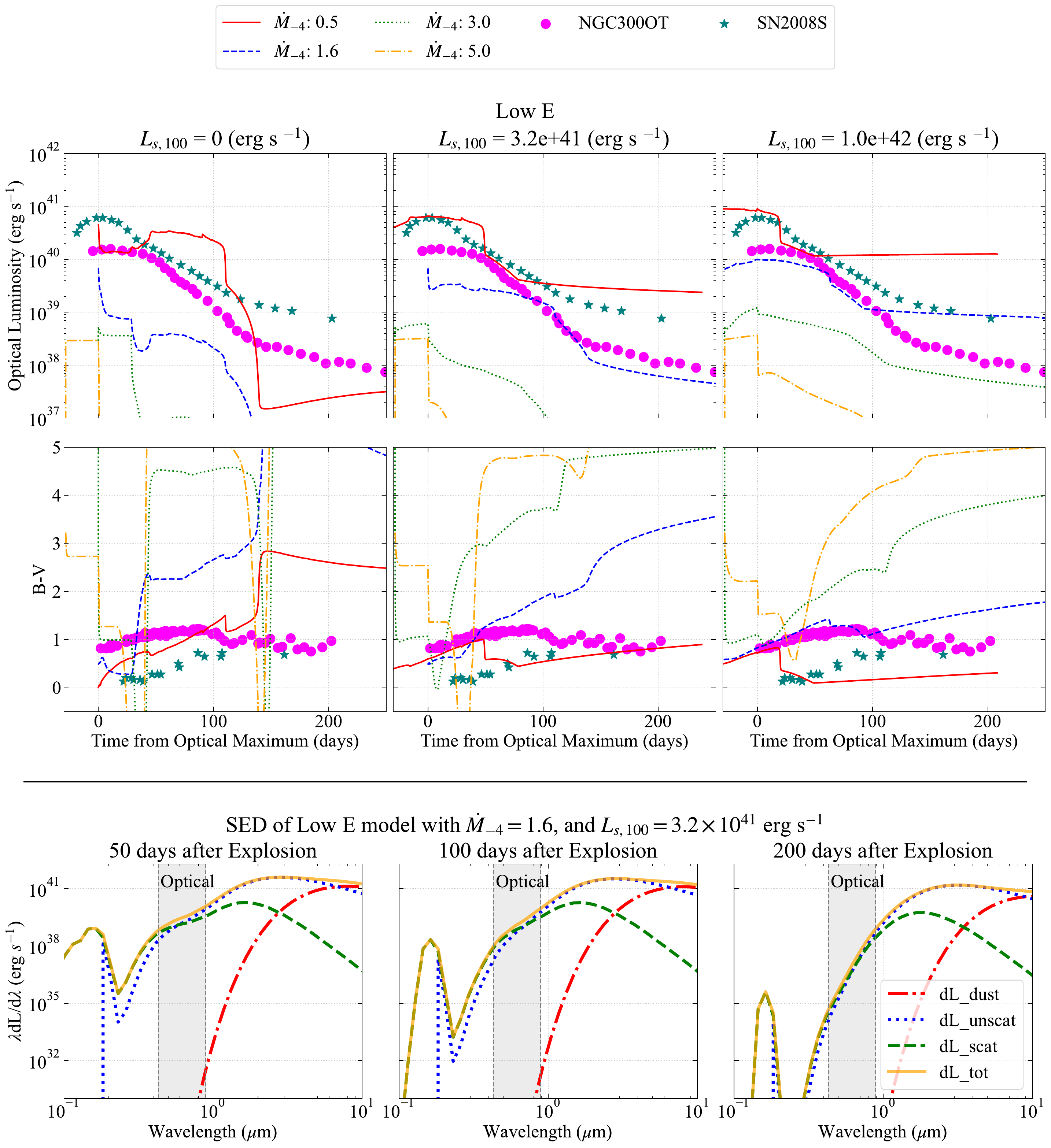}
    \caption{Same as Figure \ref{fig:LCColor_lowE} (Low E model), but including the additional attenuation by the dust surviving shock breakout. There is a strong oscillation of $B-V$ for $L_{\rm s,100}=0$ with high $\dot{M}$, which we find to be due to the rapid drop in luminosity during the transition from plateau to the tail phase. Right after the luminosity decline, the luminosity of the dust re-emission component drops as the SED peak shifts further in the IR, while the luminosity of the scattered component remains relatively unchanged since it is smoothed by the dust echo. Therefore, as long as there remains a smoothed-out component of scattered luminosity at the plateau phase, the scattered component is dominant in the optical over the dust re-emission, making the color much bluer. However, once the smoothed-out component vanishes, dust re-emission becomes dominant in the optical resulting in the drastic reddening. While this comes from our box-shape treatment of light travel effects (equation \ref{eq:light_travel}), this effect is significant only when the optical luminosity is too dim to be consistent with the ILRTs anyway.}
    \label{fig:SurvivingDust_lowE}
\end{figure*}
\begin{figure*}[t]
    \centering
    \includegraphics[width=0.9\textwidth]{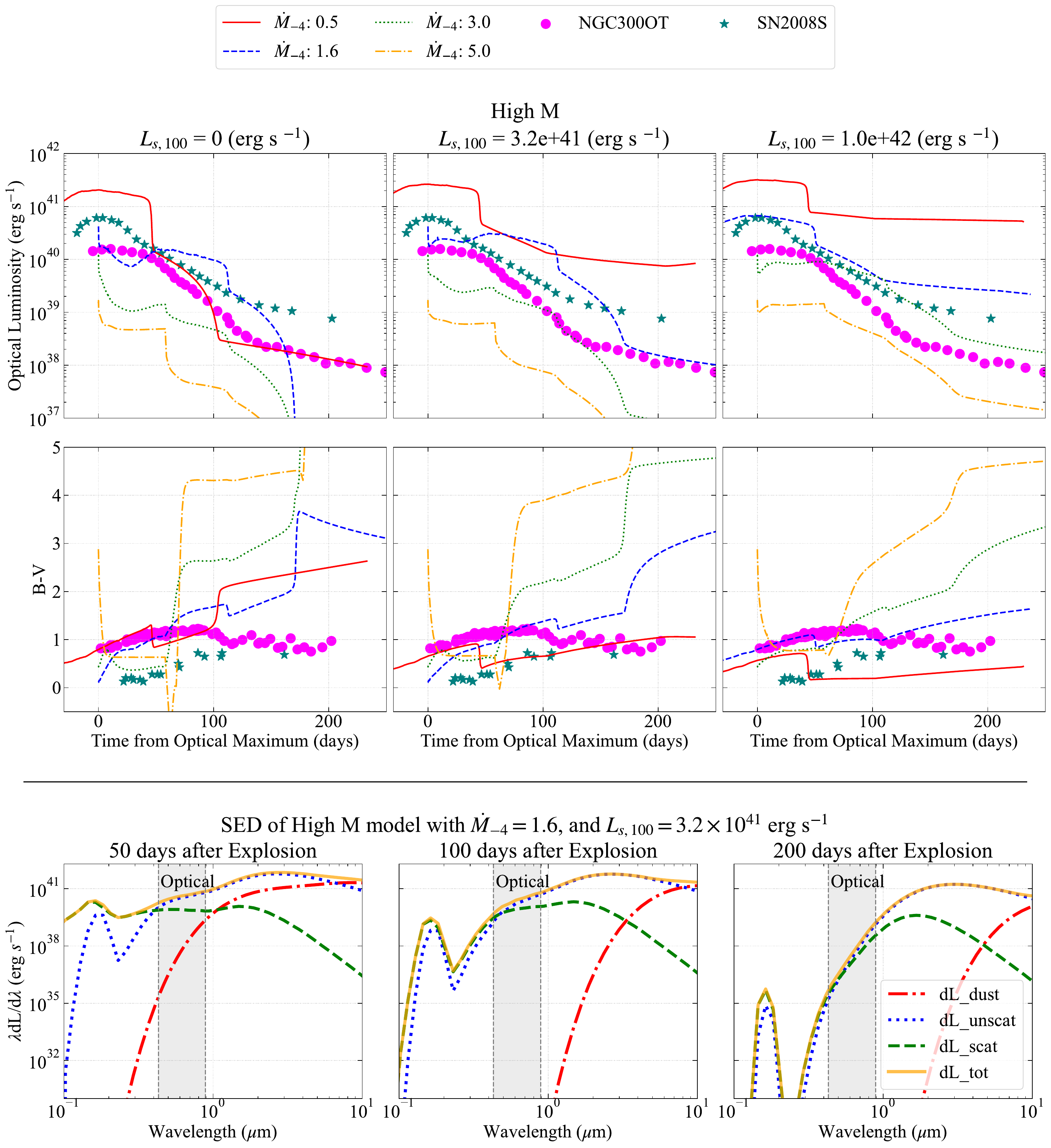}
    \caption{Same as Figure \ref{fig:LCColor_highM} (High M model), but including the additional attenuation by the dust surviving shock breakout.}
    \label{fig:SurvivingDust_highM}
\end{figure*}
\end{appendix}

\bibliography{oja_template}
\bibliographystyle{aasjournal}

\end{document}